\documentclass[letterpaper,english,aps,prx,twocolumn,amsfonts, amssymb]{revtex4}
\usepackage[T1]{fontenc}
\usepackage[utf8]{inputenc}
\usepackage{textcomp}
\usepackage{mathrsfs}
\usepackage{amsmath}
\usepackage{amssymb}
\usepackage{amsfonts}
\usepackage{makeidx}
\usepackage{graphicx}
\usepackage{esint}

\makeatletter
\usepackage[bookmarks=true,colorlinks,linkcolor=blue,urlcolor=blue,citecolor=blue]{hyperref}

\newcommand{\be}{\begin{equation}}
\newcommand{\ee}{\end{equation}}
\newcommand{\bea}{\begin{eqnarray}}
\newcommand{\eea}{\end{eqnarray}}

\newcommand{\mc}{\mathcal}

\usepackage{epsfig,graphicx,psfrag,amsmath,amssymb,float}
\usepackage[caption=false]{subfig}
\@ifundefined{showcaptionsetup}{}{%
 \PassOptionsToPackage{caption=false}{subfig}}
\usepackage{subfig}
\makeatother
\usepackage{babel}
\usepackage{float}
\begin{document}

\title{Kondo effect in a parity-time-symmetric non-Hermitian Hamiltonian}

\author{Jos\'e A. S. Louren\c{c}o}

\author{Ronivon L. Eneias}

\author{Rodrigo G. Pereira}
 
\affiliation{International Institute of Physics and Departamento de F\'isica Te\'orica
e Experimental, Universidade Federal do Rio Grande do Norte, Campus Universit\'ario, Lagoa Nova, 
Natal, RN, 59078-970, Brazil}

\begin{abstract}
The combination of non-Hermitian physics and strong correlations can give rise to new effects in open quantum many-body systems with balanced gain and loss.  We propose a generalized Anderson impurity model that includes non-Hermitian hopping terms between an embedded  quantum dot and two wires. These non-Hermitian hopping terms respect a parity-time ($\mc{PT}$) symmetry. In the   regime of a singly occupied localized state, we map the problem to a $\mc{PT}$-symmetric Kondo model and study the effects of the interactions using a perturbative renormalization group approach. We find that the Kondo effect persists if the couplings are below a critical value that corresponds to an exceptional point of the non-Hermitian Kondo interaction. On the other hand, in the regime of spontaneously broken $\mc{PT}$ symmetry, the Kondo effect is suppressed and the low-energy properties are  governed by a local-moment fixed point with vanishing conductance. 

\end{abstract}
\maketitle


\section{Introduction}

According to the fundamental postulates of quantum mechanics, any physical observable  must   be represented  by a Hermitian operator. Hermiticity  ensures that all the eigenvalues of the operator are real and therefore measurable. In particular, a Hermitean Hamiltonian guarantees  the conservation of probability in the dynamics. However, non-Hermitian Hamiltonians are routinely  used  as an approximation to describe the non-unitary dynamics of  open quantum systems \cite{moiseyev2011non,Rotter}. For instance, the imaginary part of the energy eigenvalues  can account for the decay of particles such as photons in quantum optics \cite{Plenio1998} or signal instabilities such as    the vortex  depinning transition in  superconductors \cite{Hatano1997}.  Moreover,   hermiticity is not a necessary condition for the energy spectrum to be real \cite{Bender}. Non-Hermitian Hamiltonians that preserve the symmetry composed of parity ($\mathcal{P}$) and time reversal ($\mathcal{T}$), the so-called parity-time ($\mathcal{PT}$) symmetry, can exhibit   entirely real spectra. In fact, by varying the parameters of the non-Hermitian terms in the Hamiltonian, one can find critical values (called exceptional points  \cite{Heiss2012}) at which the spectrum becomes complex. This is referred to as  spontaneous breaking of $\mc{PT}$  symmetry  \cite{Bender2} because the eigenstates of the Hamiltonian with complex eigenvalues are not eigenstates of $\mc{PT}$. 

Recent studies  of non-Hermitian Hamiltonians with $\mc{PT}$ symmetry have been stimulated  by   experiments that realize such models in open systems with balanced gain and loss \cite{Feng2017,Ramy2018,Longhireview}. Examples include optical waveguides   \cite{Guo,Ruter,Ramy2018}, cold-atomic systems  \cite{Zhang},   coupled resonators  \cite{Peng2014},    acoustic waves \cite{Shi2016,Auregan2017}, and circuit-QED \cite{Quijandria2018}. In the context of quantum many-body systems, several non-Hermitian spin chain models have been studied \cite{Korff2007,Olalla2009,Deguchi2009}.  
 It has also been proposed that non-Hermitian Hamiltonians can give rise to new topological phases with unconventional edge states \cite{Shen2018,Kawabata,Yao2018,Li2018,Gong2018}. Another intriguing possibility is the $\mc{PT}$-symmetric generalization of effective field theories \cite{BenderPRD} and quantum critical phenomena \cite{Ashida}. In Ref. \cite{Ashida}, Ashida {\it et al.} studied a $\mc{PT}$-symmetric sine-Gordon model which   describes the transition between a Tomonaga-Luttinger (TL) liquid and a Mott insulator of ultracold bosonic atoms in a one-dimensional  optical lattice with a local gain-loss structure. Remarkably, they showed that  the critical TL phase is favored by  the non-Hermitian coupling, and the insulating  phase is completely suppressed in the regime of spontaneously broken $\mc{PT}$ symmetry. 

In this work, we extend the study of $\mc{PT}$-symmetric non-Hermitian Hamiltonians to the realm of boundary critical phenomena, i.e., quantum impurity models \cite{Affleck2008}.  A paradigmatic example is the Anderson impurity model \cite{Anderson}, which   has been extensively applied to study   charge transport through semiconductor  quantum dots \cite{Pustilnik2004}. In the Coulomb-blockade regime    where  charge fluctuations can be neglected and a single electron is localized in the dot, the Anderson model can be mapped to the Kondo model via a Schrieffer-Wolff  transformation   \cite{Kondo,hewson_1993}. At low temperatures, the system exhibits the Kondo effect, whereby the effective exchange coupling grows with decreasing temperature and the  magnetic moment of the impurity gets screened via the formation of a singlet with a conduction electron. A hallmark  of the Kondo effect in quantum dots is the observation of ideal quantized conductance at low temperatures \cite{GoldhaberGordon1998,Cronenwett1998}.

Here  we propose a generalization of the Anderson impurity model in which an embedded quantum dot is weakly coupled to two leads by non-Hermitian hopping terms. The latter can be engineered by means of auxiliary sites with complex potentials \cite{Longhi}. Performing a Schrieffer-Wolff  transformation, we obtain a $\mc{PT}$-symmetric non-Hermitian Kondo model. We analyze the effects of the Kondo interactions using the perturbative renormalization group (RG) \cite{Anderson1970,Wilson1975,Affleck:1995}. We find two regimes, depending on the ratio $g$ between the coupling of the  non-Hermitian  term and the conventional Kondo coupling. For $g<1$,  the Kondo effect persists and the system flows to strong coupling at low energies. Analyzing    the local tight-binding model at strong coupling, we find that in this regime the spectrum  is real and  the formation of the Kondo singlet with a $\mc{PT}$-symmetric orbital  leads to a stable fixed point with ideal conductance and emergent $\mc P$ and $\mc T$ symmetries. On the other hand, for $g>1$, the spectrum becomes complex and the $\mc{PT}$ symmetry is spontaneously broken. However, in this regime the perturbative RG flow is towards a local-moment fixed point \cite{hewson_1993}, in which the impurity spin decouples from the leads and the conductance vanishes. Therefore, the Kondo effect is suppressed in the broken-$\mc{PT}$ regime. Our model can in principle be implemented experimentally by means of   two-terminal transport measurements in cold atomic gases  \cite{Chien,Krinner2017} with controlled  loss and gain (the latter being achieved by  pumping atoms into the auxiliary sites \cite{Robins2008}).


The   paper is organized as follows. First, in Sec. \ref{Model}, we introduce  the  Anderson impurity model with  $\mathcal{PT}$-symmetric non-Hermitian hopping  between the localized state and the wires. We also discuss the mapping to the $\mathcal{PT}$-symmetric Kondo model. Next,   in Sec. \ref{RG}, we take the continuum limit and derive  the  RG equations  for the Kondo couplings, identifying   two distinct regimes in the flow diagrams  as a function of the dimensionless parameter   $g$. In Sec. \ref{SCS}, we investigate the  spectrum in the strong coupling limit and relate the spontaneous breaking of $\mc{PT}$ symmetry to the absence of the Kondo effect. The observable effects on the conductance through the quantum dot are discussed in Sec. \ref{Transmittance}. We finally close with the conclusions in Sec. \ref{Conclusions}. 

\section {Model} \label{Model}

\begin{figure}[t]
	\centering
	\includegraphics[width=0.85\columnwidth]{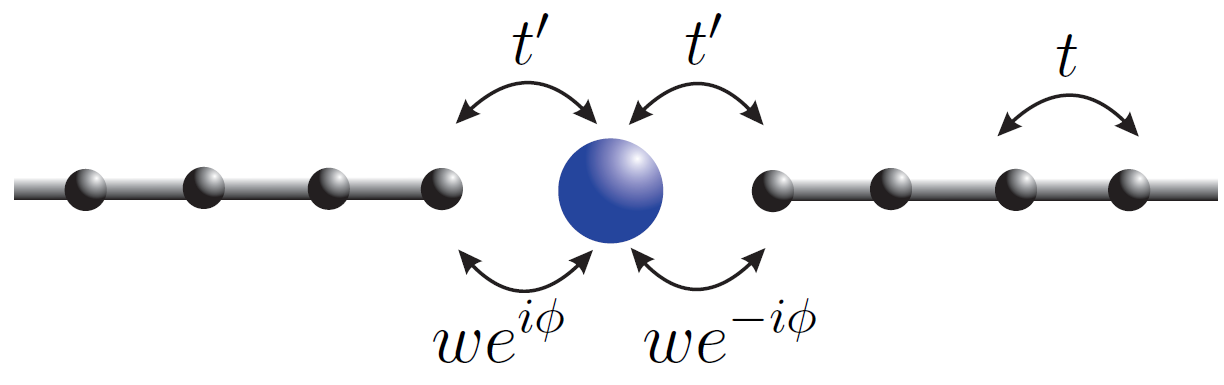}\\
	\caption{Schematic representation of the Anderson model with  $\mathcal{PT}$-symmetric non-Hermitian hopping between the quantum dot and the wires.}
	\label{fig.1}
\end{figure}

We study the Anderson model that describes the transport between two wires across a quantum dot as illustrated in Fig. \ref{fig.1}. In addition to the usual direct  hopping  between the dot and the   wires, we consider a non-Hermitian hopping  process  which represents an alternative tunnelling path through auxiliary sites   coupled  to a particle reservoir  \cite{Longhi}.  The Hamiltonian is 
\begin{eqnarray}\label{eq1}\
H&=&H_{0}+H_{t'}+H_{d}+H_{\mathcal{PT}},\\\nonumber\label{eq2}
H_0&=&-t{\displaystyle\sum_{j\leq-2}}\left(c_{j}^{\dagger}c^{\phantom\dagger}_{j+1}+\text{h.c.}\right)\\
 &&-t{\displaystyle\sum_{j\geq1}}\left(c_{j}^{\dagger}c^{\phantom\dagger}_{j+1}+\text{h.c.}\right),\\\label{eq3}
H_{t'}&=&-t'\left[c_d^{\dagger}\left(c^{\phantom\dagger}_{-1}+c^{\phantom\dagger}_{1}\right)+\text{h.c.}\right],\\\label{eq4}
H_d&=&\epsilon_{d}c_d^{\dagger}c^{\phantom\dagger}_d+Un_{d\uparrow}n_{d\downarrow},\\\nonumber\label{eq5}
H_{\mathcal{PT}}&=&w{\rm e}^{i\phi}\left(c_{1}^{\dagger}c^{\phantom\dagger}_d+c_d^{\dagger}c^{\phantom\dagger}_{1}\right)\\
&&+w{\rm e}^{-i\phi}\left(c_{-1}^{\dagger}c_d+c_d^{\dagger}c^{\phantom\dagger}_{-1}\right),
\end{eqnarray}
where $c_d=\left(c_{d\uparrow},c_{d\downarrow}\right)^{T}$ is the two-component spinor of annihilation operators of   electrons (or spin-$1/2$ fermionic atoms in cold-atom realizations \cite{Esslinger2010,Riegger2018}) in the localized state of the quantum dot, $c_j=\left(c_{j\uparrow},c_{j\downarrow}\right)^{T}$ represents the  states in the wires (with $c_j$ acting in the left wire for $j\leq -1$ or in the right wire for $j\geq 1$), $t$ is the hopping parameter   in the wires, $t'$ is the amplitude for hopping between the localized state and the ends of wires, and $w{\rm e}^{i\phi}$ ($w{\rm e}^{-i\phi}$), with $w\in\mathbb R$ and $\phi\in[-\pi,\pi]$, is the complex  hopping amplitude between the localized state and the wire on  the left  (right). The Hamiltonian is non-Hermitian for $\phi \neq 0, \pi$, but preserves $\mc{PT}$  symmetry with parity and time reversal transformations defined by \bea
\mc P:&\;&c_{j}\mapsto c_{-j},\\
\mc T:&\;&i\mapsto -i,\,c_{j}\mapsto i\sigma^yc_{j},\,c_{d}\mapsto i\sigma^yc_{d},
\eea
where $\sigma^y$ is the   Pauli matrix in spin space. In the dot Hamiltonian $H_d$, $\epsilon_{d}<0$ is the energy of an electron  in the localized state, $n_{d\sigma}=c^\dagger_{d\sigma}c^{\phantom\dagger}_{d\sigma}$ is the number operator for  spin $\sigma=\uparrow, \downarrow$,  and $U>0$ is the repulsive interaction strength  between two electrons in the dot. The model is particle-hole symmetric for $U=-2\epsilon_d$    and   Fermi momentum $k_F=\pi \langle c^\dagger_jc^{\phantom\dagger}_j\rangle/2=\pi/2$ in the wires (setting the lattice spacing $a=1$). In this work, we shall be mainly interested in the particle-hole symmetric case.

The Hamiltonian in Eq. (\ref{eq1}) can be mapped to a Kondo model in the regime $t',w\ll |\epsilon_{d}|, U$ \cite{hewson_1993}. In this case, we consider that in the low-energy subspace the localized  state is occupied by a single electron  with spin $\uparrow$ or $\downarrow$ (such that $n_d=\sum_{\sigma}n_{d\sigma}=1$).  The Schrieffer-Wolff transformation generates an effective spin exchange interaction in the low-energy subspace by projecting out   the high-energy  states with $n_d=0$ or $n_d=2$. To second-order perturbation theory, we obtain the effective Hamiltonian  $H_{\rm eff}=H_0+H_I$, where the Kondo interaction $H_I$ has the form 
\begin{eqnarray}\label{eq6}
H_I&=&J_0\,{\displaystyle\left(c_{1}^{\dagger}+c_{-1}^{\dagger}\right)\frac{\boldsymbol{\sigma}}{2}\left(c^{\phantom\dagger}_{1}+c^{\phantom\dagger}_{-1}\right)\cdotp\mathbf{S}}\nonumber\\ \nonumber
&&{\displaystyle-iJ_1\left(c_{-1}^{\dagger}\frac{\boldsymbol{\sigma}}{2}c^{\phantom\dagger}_{-1}-c_{1}^{\dagger}\frac{\boldsymbol{\sigma}}{2}c_{1}\right)\cdotp\mathbf{S}}\\ 
&&-J_2\left(c_{1}^{\dagger}\frac{\boldsymbol{\sigma}}{2}c_{1}+c_{-1}^{\dagger}\frac{\boldsymbol{\sigma}}{2}c^{\phantom\dagger}_{-1}\right)\cdotp\mathbf{S}.
\end{eqnarray}
Here $\boldsymbol{\sigma}$ denotes the vector of Pauli matrices and $\mathbf S$ is the spin-$1/2$ operator of the localized electron. The exchange coupling constants are given by \bea
J_{0}&=& J+ J'+2\sqrt{JJ'} \cos\phi,\label{Jcoupling1}\\
 J_{1}&=&2\sqrt{JJ'}\sin\phi+J' \sin\left(2\phi\right) ,\label{Jcoupling2}\\
   J_{2}&=& J'\,\left[1-\cos\left(2\phi\right)\right]\label{Jcoupling3},
\eea
where
\begin{eqnarray}\label{eq9}
J&=&2{t'}^2 \left(\frac{1}{-\epsilon_{d}}+\frac{1}{U+\epsilon_{d}}\right) \ , \\\label{eq10}
J'&=& 2{w}^2 \left(\frac{1}{-\epsilon_{d}}+\frac{1}{U+\epsilon_{d}}\right) \ .
\end{eqnarray}
Note that $J_0,J_2\geq 0$. The first term in Eq. (\ref{eq6}) corresponds to the standard antiferromagnetic Kondo coupling between the impurity spin and the symmetric orbital on sites $j=1$ and $j=-1$ of the  wires \cite{Simon}. This is the only coupling that survives in the Hermitian case $\phi=0,\pi$. By contrast, $J_2$ represents a ferromagnetic two-channel Kondo coupling \cite{Nozieresph,Zawadowski} between the impurity and the   spins  at the ends of the wires. Finally, $J_1$ is the coupling constant of the   non-Hermitian term. This term is odd under both $\mc P$ and $\mc T$ (with $\mc T:\mathbf S\mapsto -\mathbf S$ for the impurity spin), thus preserving the $\mc{PT}$ symmetry of the original   model in Eq. (\ref{eq1}).  
 
 
\section{Renormalization Group}\label{RG}
The Kondo effect can be understood within a perturbative  RG analysis  \cite{Anderson1970,hewson_1993,Affleck:1995}. First, let us recall the result for the conventional Kondo model for an embedded quantum dot, which  corresponds to setting $J_1=J_2=0$ in Eq. (\ref{eq6}).  In the RG analysis, the constant $J_0$ must be replaced by an effective interaction $J_0(\Lambda)$ that depends on the energy scale $\Lambda$ at  which the properties of the system are measured. In the low-energy limit, $\Lambda\to 0$, the effective interaction diverges. The interpretation is that the localized spin forms a singlet with an electron in the symmetric channel between the two wires. The  low-energy physics is described by a Fermi liquid fixed point  \cite{Nozieres1974} at which  the boundary conditions on conduction  electrons are modified by a universal phase shift (in the case of particle-hole symmetry),  leading  to an ideal conductance between the two wires \cite{Glazman2003}. 

\begin{figure}[t]
	\centering
	\includegraphics[width=0.9\columnwidth]{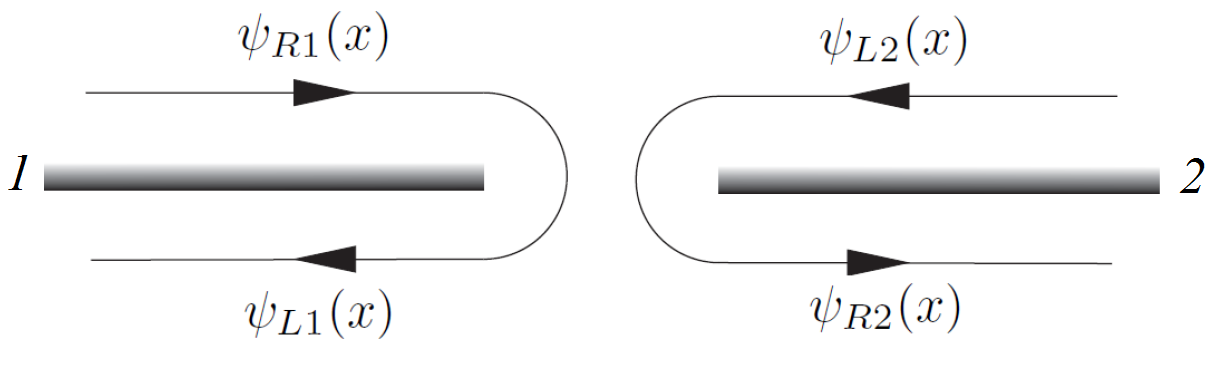}\\
	\caption{Low-energy   modes for electrons in the  two      wires with open boundary conditions at $x=0$.}
	\label{fig.2}
\end{figure}

We now consider the $\mc{PT}$-symmetric  Kondo model    in the weak coupling regime  $J_0,J_1,J_2\ll t$. For $J_0=J_1=J_2=0$, the Hamiltonian in Eq.  (\ref{eq6}) describes two decoupled tight-binding models with open boundary conditions at $j=0$. This free Hamiltonian can be diagonalized using  the Fourier transform\bea
c_{j<0}=\int_0^\pi\frac{dk}{\pi} \sin(kj)c_{k1},\\
c_{j>0}=\int_0^\pi\frac{dk}{\pi} \sin(kj)c_{k2},
\eea
where $c_{k\eta}$,   with $\eta=1,2$, are the annihilation operators of electrons with momentum $k$  in  the  wire on the left  for $\eta=1$ or   on the right for $\eta=2$. The operators   $c_{k\eta}$ obey $\{c^{\phantom\dagger}_{k\eta},c^\dagger_{k'\eta'}\}=2\pi\delta_{\eta\eta'}\delta(k-k')$. We can then write\be
H_0=\sum_{\eta=1,2}\int_0^\pi \frac{dk}{2\pi} \varepsilon(k)c^\dagger_{k\eta}c^{\phantom\dagger}_{k\eta},
\ee
where $\varepsilon(k)=-2t\cos(k)$ is the dispersion relation. At half-filling (the particle-hole symmetric case), the ground state is constructed by occupying  the single-particle states with $0<k<k_F=\pi/2$. We  take  the continuum  limit  by linearizing the spectrum around the  Fermi point. In real space, the operators $c_j$ are replaced by the fields  $\psi_\eta(x)$  in the form \cite{Simon}
\begin{eqnarray}\label{eq12}
c_{j<0}\rightarrow\psi_1(x=j)\sim e^{ik_{F}x}\psi_{R1}(x)+e^{-ik_{F}x}\psi_{L1}(x),\nonumber\\
c_{j>0}\rightarrow\psi_2(x=j)\sim e^{ik_{F}x}\psi_{R2}(x)+e^{-ik_{F}x}\psi_{L2}(x).\label{modes}
\end{eqnarray}
Here $\psi_{R/L,\eta} (x)$ are the slowly varying right- or left-moving components of the fermionic field, respectively. We impose open boundary conditions, $\psi_\eta(0)=0$, by the relation\be
\psi_{L\eta}(x)=-\psi_{R\eta}(-x).\label{openbc}
\ee
Thus, the left movers in wire $\eta=1$ (the outgoing modes with respect to scattering at the boundary) can be regarded as the analytic continuation of the right movers to the positive-$x$ axis (see Fig. \ref{fig.2}). Likewise, we treat  the right movers in wire $\eta=2$  as the analytic continuation of the left movers. Defining the four-component spinor
\begin{eqnarray}\label{eq14}
\Psi(x)=\left(\begin{array}{cc} 
\psi_{L2}(-x) \\ \psi_{R1}(x) \end{array}\right)=\left(\begin{array}{cccc} 
\psi_{L2 \uparrow}(-x) \\ \psi_{L2 \downarrow}(-x) \\ \psi_{R1 \uparrow}(x) \\  \psi_{R1 \downarrow}(x)\end{array}\right),
\end{eqnarray}
we can write the free Hamiltonian in the low-energy approximation as \begin{eqnarray}\label{eq13}
 H_{0}&\approx &v_{F}\int_{-\infty}^{+\infty}dx\Psi^{\dagger}(x)(-i\partial_x)\Psi(x),
\end{eqnarray}
where $v_F = 2t \sin (k_F)$ is the Fermi velocity.

We now rewrite the interacting part of the Hamiltonian   in the continuum limit using Eqs. (\ref{modes}) and (\ref{openbc}) with $\psi_{R/L,\eta}(\pm 1)\approx\psi_{R/L,\eta}(0)$. The result is 
\be\label{eq15}
 H_{I}\approx \pi v_F\Psi^\dagger(0)\left(\lambda_0\boldsymbol{\Sigma}+i\lambda_1\boldsymbol{\Omega}-\lambda_2\boldsymbol{\Gamma}\right)\Psi(0)\cdot \mathbf S,
\ee
where $\displaystyle \lambda_n= J_n \sin(k_F)/(\pi t)$ with $n=0,1,2$ are the   dimensionless Kondo couplings and the $4\times4$  matrices $\boldsymbol{\Sigma}$, $\boldsymbol{\Omega}$ and $\boldsymbol{\Gamma}$ are written in terms of the Pauli matrices as follows:
\begin{eqnarray}\label{eq16}
 \boldsymbol{\Sigma}=\left(\begin{array}{cc}
\boldsymbol{\sigma} & \boldsymbol{\sigma}\\
\boldsymbol{\sigma} & \boldsymbol{\sigma}
\end{array}\right), \,
\boldsymbol{\Omega}=\left(\begin{array}{cc}
\boldsymbol{\sigma} & 0\\
0 & -\boldsymbol{\sigma}
\end{array}\right),\, 
\boldsymbol{\Gamma}=\left(\begin{array}{cc}
\boldsymbol{\sigma} & 0\\
0 & \boldsymbol{\sigma}
\end{array}\right).
\end{eqnarray}
These matrices obey the  algebra
\begin{eqnarray}\nonumber\label{eq17}
 \left[\Sigma^{a},\Sigma^{b}\right]&=&2\left[\Sigma^{a},\Gamma^{b}\right]=2\left[\Gamma^{a},\Sigma^{b}\right]=4i\epsilon^{abc}\Sigma^{c},\\ \nonumber
\left[\Gamma^{a},\Gamma^{b}\right]&=&\left[\Omega^{a},\Omega^{b}\right]=2i\epsilon^{abc}\Gamma^{c},\\  \nonumber
\left[\Omega^{a},\Gamma^{b}\right]&=&2i\epsilon^{abc}\Omega^{c},\\
\left[\Omega^{a},\Sigma^{b}\right]&=&2i\epsilon^{abc}\Omega^{c}+2i\delta^{ab}\Upsilon, 
\end{eqnarray}
where $\Upsilon=\left(\begin{array}{cc}
0 & -i\mathbb I_2\\
i\mathbb I_2 & 0
\end{array}\right)$ and  $\mathbb I_2$ is the $2\times2$ identity matrix. Equation (\ref{eq15})  contains the most general boundary exchange interactions allowed by (spin-rotation) SU(2) and $\mc{PT}$ symmetries.

We calculate the RG equations  using perturbation theory  to second order in couplings $\lambda_0$, $\lambda_1$ and $\lambda_2$ following the procedure for the   Kondo model  \cite{Affleck:1995}. In this calculation, we employ  the algebra in Eq. (\ref{eq17}). We also use the time-ordered matrix Green's function for free electrons \begin{eqnarray}\label{eq18}
G(t-t')&=& -i\left\langle T\Psi(0,t)\Psi^{\dagger}(0,t')\right\rangle\nonumber\\
&=& -\frac{\mathbb{I}_4}{2\pi v_F(t-t')},
\end{eqnarray}
where $\mathbb I_4$ is the $4\times4$ identity matrix.  
In the RG step, we integrate out short time intervals between scattering processes, $\Lambda^{-1}<|t-t'|<(\Lambda')^{-1}$, where   $\Lambda$ and $\Lambda'=\Lambda-d\Lambda$ are the old and new high-energy  cutoff scales, respectively. 
We obtain  the followings set of RG  equations:
\begin{eqnarray}\label{eq19}
\frac{d\lambda_{0}}{d\ell}&=&\lambda_{0}^{2}-\lambda_{0}\lambda_{2}\ ,\\ \label{eq20}
\frac{d\lambda_{1}}{d\ell}&=&\lambda_{0}\lambda_{1}-\lambda_{1}\lambda_{2}\ ,\\\label{eq21}
\frac{d\lambda_{2}}{d\ell}&=&-\frac{\lambda_{2}^{2}}{2}+\frac{\lambda_{1}^{2}}{2},
\end{eqnarray} 
where $d\ell=d\Lambda/\Lambda$. At first sight, these   RG equations involve three independent couplings, generating a  three-dimensional   flow diagram. However,   by combining Eqs.  (\ref{eq19}) and (\ref{eq20}), one can verify that  the ratio
\begin{eqnarray}\label{eq22}
g\equiv\frac{\lambda_{1}}{\lambda_{0}}
\end{eqnarray}
is conserved along the RG flow, i.e., $dg/d\ell=0$,  at least for the beta functions calculated to second order in the couplings. Substituting $J_1=gJ_0$, we are left with only two coupled RG equations:  \begin{eqnarray}\label{copyeq19}
\frac{d\lambda_{0}}{d\ell}&=&\lambda_{0}^{2}-\lambda_{0}\lambda_{2} , \\ \label{eq23}
\frac{d\lambda_{2}}{d\ell}&=&-\frac{{\lambda}_2^2}{2}+g^2\frac{\lambda_{0}^2}{2}.
\end{eqnarray}

\begin{figure}[t]
	\subfloat[]{
	\includegraphics[height=4cm]{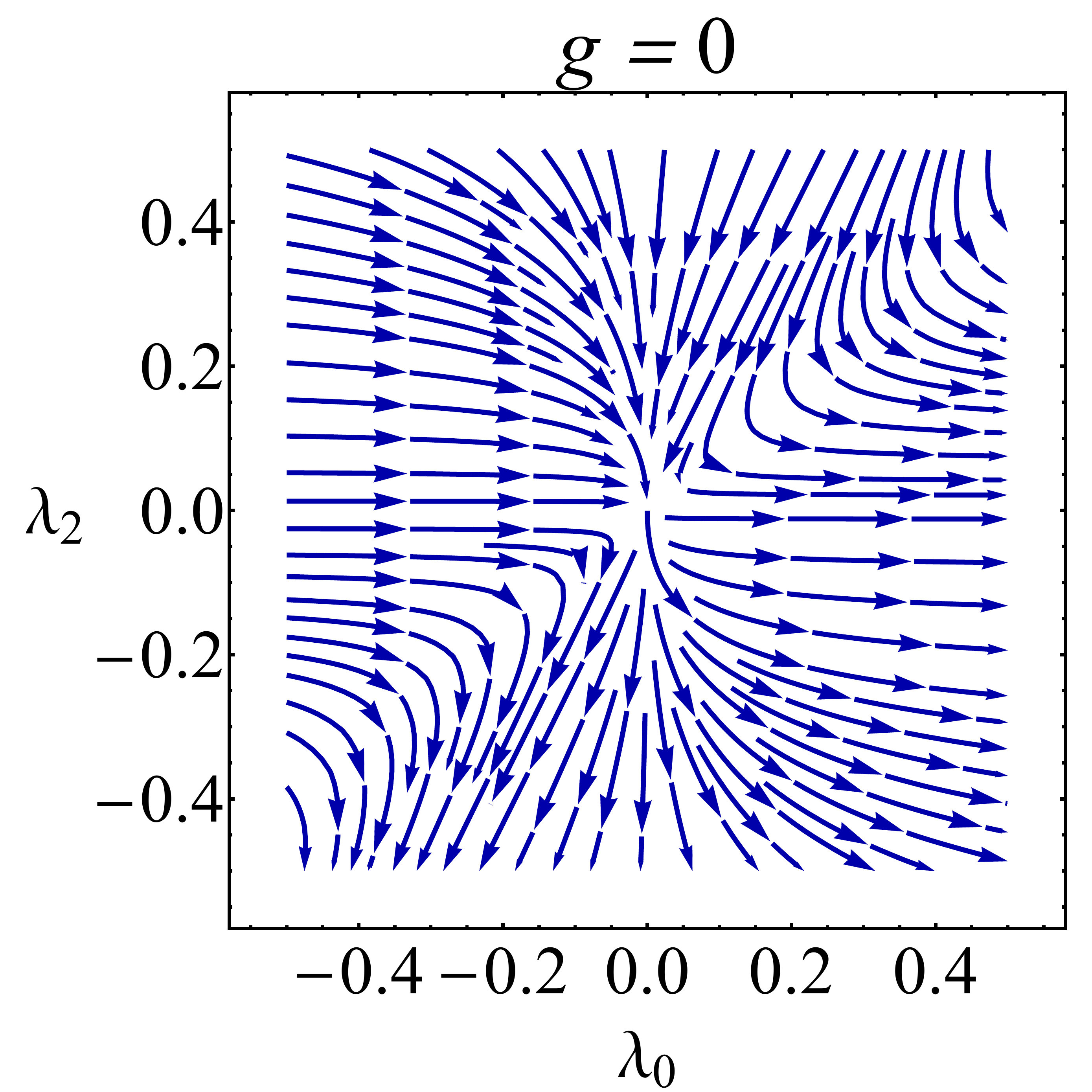}}
	\subfloat[]{
    \includegraphics[height=4cm]{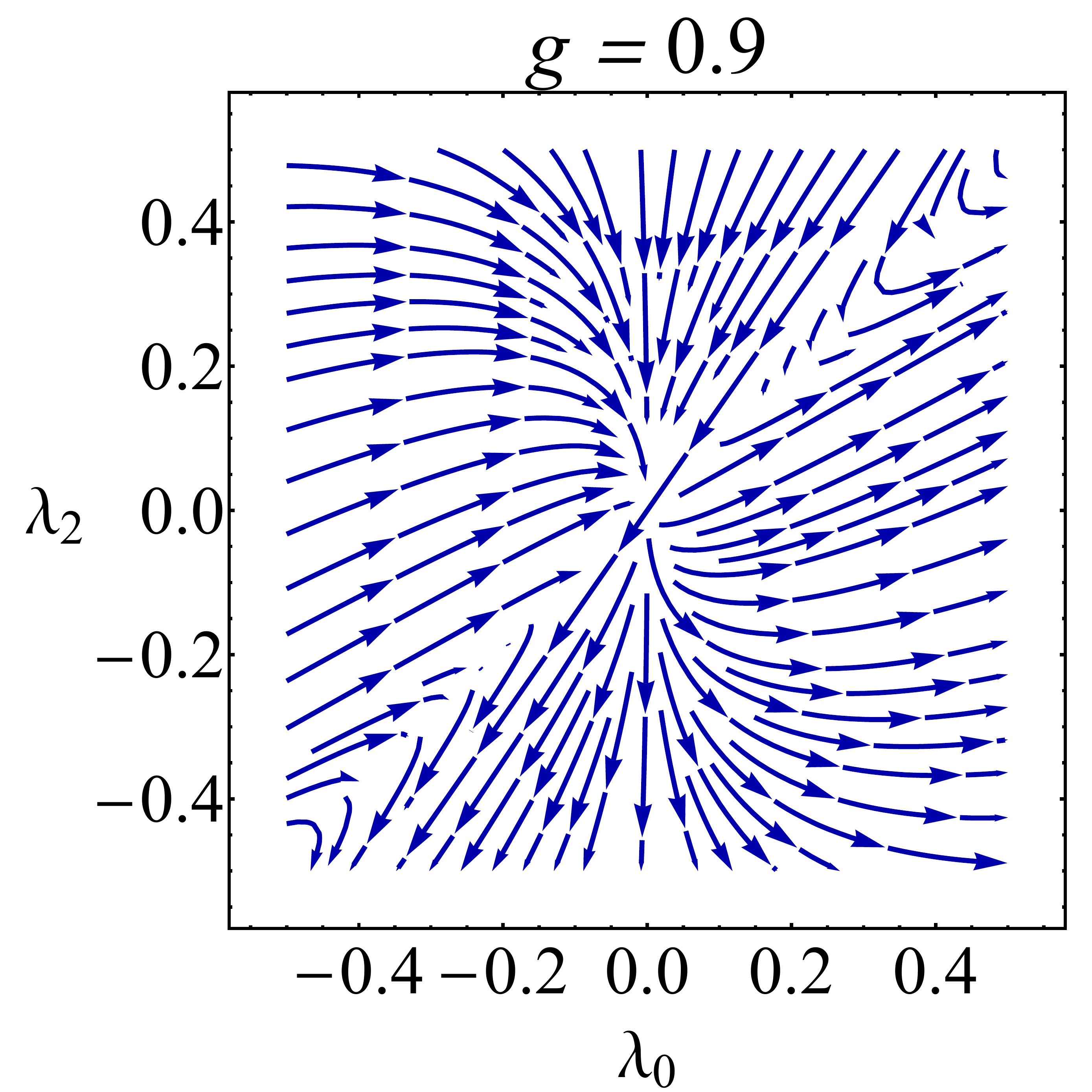}}\\ 
    \subfloat[]{
    \includegraphics[height=4cm]{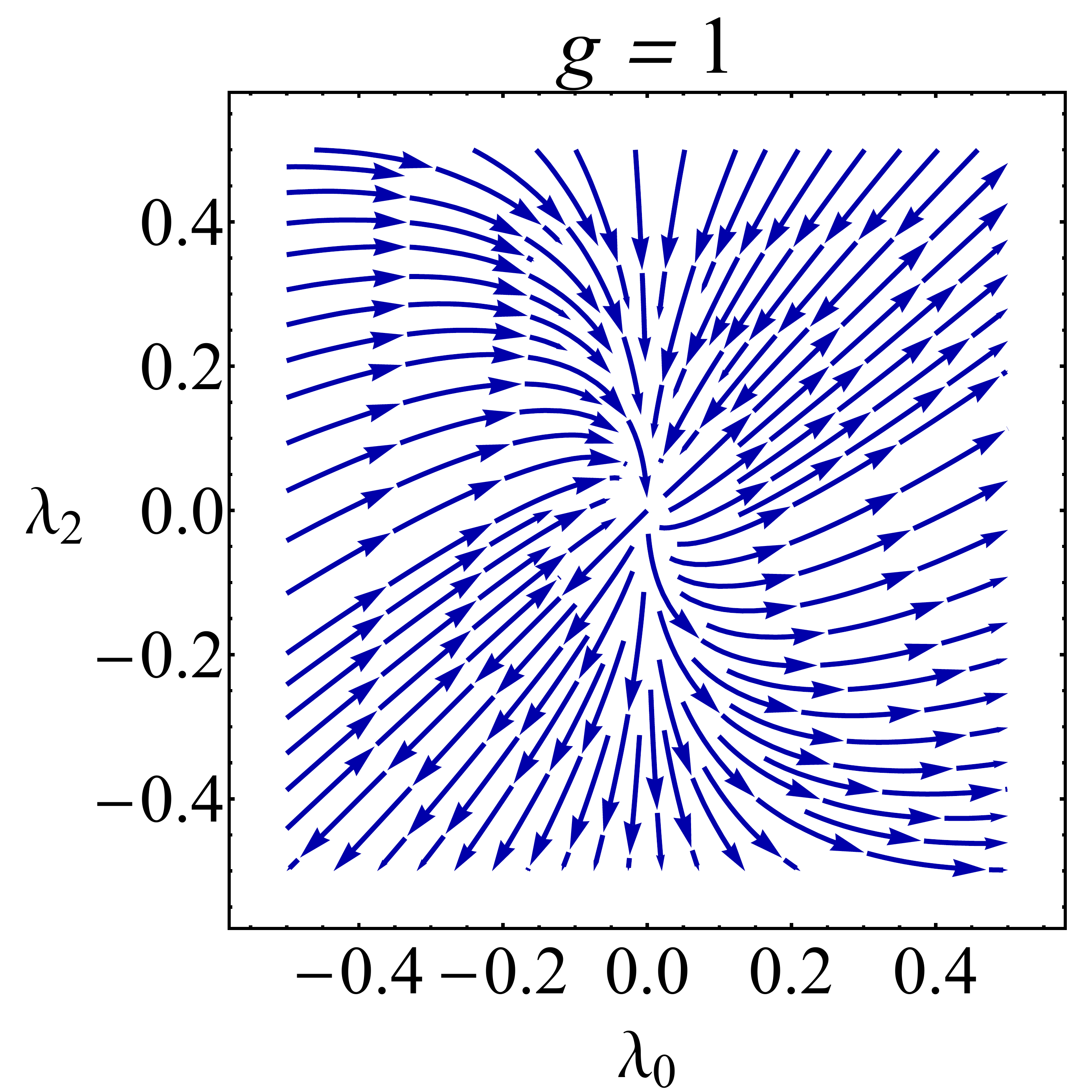}}\quad 
    \subfloat[]{
    \includegraphics[height=4cm]{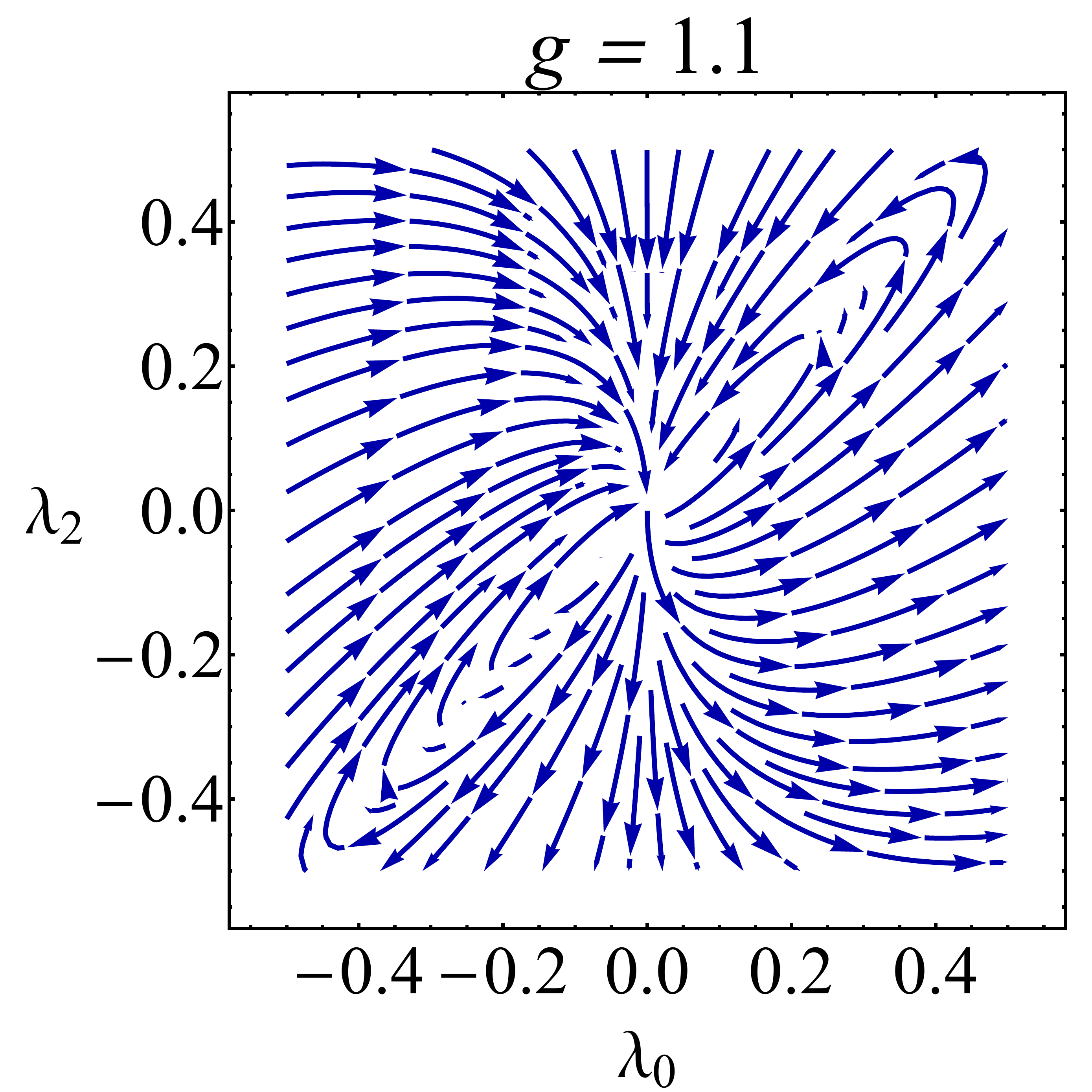}}
    \caption{RG flow diagrams for  the couplings $\lambda_{0}$ and $\lambda_{2}$ and different values of $g$: (a) $g=0$, (b) $g=0.9$, (c) $g=1$, and (d) $g=1.1$.}
	\label{fig.3}
\end{figure}

Figure \ref{fig.3}  shows the     RG flow according to Eqs.  (\ref{copyeq19}) and (\ref{eq23}) for different values of  $g$. Fig. \ref{fig.3}(a) corresponds to the Hermitian case $g=0$.    Along the line $\lambda_2=0$, we recover the usual Kondo effect \cite{hewson_1993}: the dimensionless Kondo coupling $\lambda_0$ is marginally irrelevant in the ferromagnetic case $\lambda_0<0$ and marginally relevant in the antiferromagnetic case $\lambda_0>0$. In the regime $0<\lambda_2<\lambda_0$, we have $\lambda_0(\Lambda)\to \infty$ while $\lambda_2(\Lambda)\to 0$ as $\Lambda\to0$. When we turn on $g\neq 0$, as in Fig. \ref{fig.3}(b),  the flow to strong coupling is no longer along the $\lambda_2=0$ line because $\lambda_2$ grows together with $\lambda_0$. This suggests that the presence of the non-Hermitian term affects the asymptotic value  of the ratio $\lambda_{2}/\lambda_0$ in the low-energy limit. Figure \ref{fig.3}(c) shows that $|g|=1$ is a critical value, characterized by a discontinuity of the flow across the line $\lambda_2=\lambda_0$. For $|g|>1$, as in Fig.  \ref{fig.3}(d), the flow becomes non-monotonic: for $0<\lambda_2<\lambda_0$, the couplings  initially increase, but eventually turn around and flow back to the non-interacting fixed point $\lambda_0=\lambda_2=0$, regardless of their initial values. Similar unconventional behavior is observed in the renormalization of the interactions of the non-Hermitian  sine-Gordon model in the  $\mc {PT}$-broken phase \cite{Ashida}.  

\section{Spectrum in the strong coupling limit}\label{SCS}

We saw in Sec. \ref{RG} that, for   $|g|<1$ and depending on the bare values of $\lambda_0,\lambda_2>0$, the system can flow to strong coupling in the low-energy limit. As in the usual Kondo effect, we can understand the strong-coupling fixed point  by  going back to  the lattice model and analyzing the limit in which the interactions are dominant, $J_0, J_2\gg t$ \cite{Simon}. In this limit, we start by diagonalizing $H_I$ in Eq. (\ref{eq6}). The latter can be viewed as a three-site operator that acts in the Hilbert space $\mathcal{H}=\mathcal{H}_{-1}\otimes\mathcal{H}_{0}\otimes\mathcal{H}_{1}$, where $\mathcal{H}_{\pm1}=\{|0\rangle,|\uparrow\rangle, |\downarrow\rangle, |\uparrow\downarrow\rangle,\}$ are the local Hilbert spaces of electrons in sites $j=\pm1$ and $\mc H_0=\{|\Uparrow\rangle,|\Downarrow\rangle\}$ is the Hilbert space of the impurity spin.  

We can block diagonalize the Kondo interaction $H_I$ in sectors labeled by the    total number of electrons $N_e=c^\dagger_{-1}c^{\phantom\dagger}_{-1}+c^\dagger_1c^{\phantom\dagger}_1$ and by one component of the total spin $S^z_{\text{tot}}=c^\dagger_{-1}\frac{ \sigma^z}2c^{\phantom\dagger}_{-1}+S^z+c^\dagger_1\frac{ \sigma^z}2c^{\phantom\dagger}_1$. The possible values for these good quantum numbers are $N_e=0,1,\dots, 4$ and $-(\tilde N_e+1)/2\leq S^z_{\text{tot}}\leq (\tilde N_e+1)/2$, where $\tilde N_e=\text{min}\{N_e,4-N_e\}$. We also have the selection rule that $S^z_{\text{tot}}$ is integer if $N_e$ is odd and half-integer if $N_e$ is even. Due to particle-hole symmetry, the spectrum for $N_e=n$ electrons is the same as for $N_e=4-n$ electrons. Thus, we can restrict ourselves to $0\leq N_e\leq 2$. Likewise, due to SU(2) symmetry, the spectrum for spin $S^z_{\text{tot}}=m$  is the same as for $S^z_{\text{tot}}=-m$ and we focus on $m\geq0$.  
 
Let us denote the energy levels in each subspace by $E_l(N_e,S^z_{\text{tot}})$, where $l$ runs from $l=1$ to the dimension of the subspace. In the sector with $N_e=0$, the interacting Hamiltonian  $H_I$ vanishes identically, thus $E_{l}(0,1/2)=0$, with $l=1,2$. For $N_e=1$ and $S^{z}_{\text{tot}}=1$, we find two energy levels given by  \be
E_{1,2}(1,1)=\frac{1}{4}\left(J_0-J_2\pm J_0\sqrt{1-g^2}\right). \label{E12}
\ee
Note that these energies are real for $|g|\leq1$. For $|g|>1$, $E_{1,2}(1,1)$ form  a complex conjugate pair. This corresponds to the spontaneous breaking of $\mc{PT}$ symmetry and it is a first sign that $g=\pm1$ are exceptional points of the Kondo interactions. We confirm this expectation by calculating the energy levels in the other sectors. For $N_e=1$ and $S^{z}_{\text{tot}}=0$, the eigenvalues are \bea
E_{1,2}(1,0)&=&-\frac{3}{4}\left(J_0-J_2\pm J_0\sqrt{1-g^2}\right),\\
E_{3,4}(1,0)&=&\frac{1}{4}\left(J_0-J_2\pm J_0\sqrt{1-g^2}\right). 
\eea
Once again, the energies are real for $|g|\leq1$. Here $E_{1}(1,0)$ and $E_2(1,0)$ are associated with singlet states (i.e., eigenstates of $\mathbf S_{\text{tot}}^2$ with $S_{\text{tot}}=0$). Their wave functions are    antisymmetric in the  spin part, but  they correspond to different orbital states. For $|g|<1$, the singlet state with the lowest energy, $E_{1}(1,0)$, has the form \be
|\Psi_0\rangle = \frac12\left[\left|\uparrow,\Downarrow,0\right\rangle-\left|\downarrow,\Uparrow,0\right\rangle+e^{i\alpha}\left(\left|0,\Downarrow,\uparrow\right\rangle- \left|0,\Uparrow,\downarrow\right\rangle\right)\right],\label{singlet}
\ee 
where $\alpha=\text{arctan}(g/\sqrt{1-g^2})$. Note that  $|\Psi_0\rangle$   is an eigenstate of $\mc {PT}$ and for $g=0$ it reduces to the singlet state in the symmetric orbital, where the   electron is in the superposition $(|j=-1\rangle+|j=1\rangle)/\sqrt2$. 

\begin{figure}[t]
	\centering
	\includegraphics[width=0.95\columnwidth]{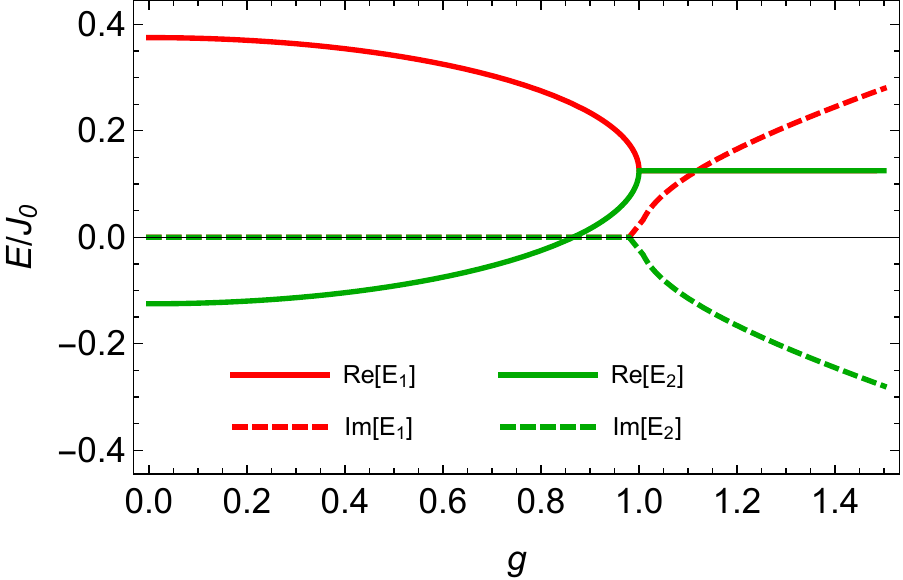} 
	\caption{Real and imaginary parts of the energies in the sector with $N_e=1$ and $S^z_{\text{tot}}=1$ as given in Eq. (\ref{E12}). Here  we set $J_2=J_0/2$.  The spectrum becomes complex for $|g|>1$.}
	\label{fig.4}
\end{figure}

For $N_e=2$ and $S^{z}_{\text{tot}}=3/2$, we have only one  state (${|\uparrow,\Uparrow,\uparrow\rangle}$) with energy $E_1(2,3/2)=(J_0-J_2)/2$, independent of the parameter $g$.  Finally, for $N_e=2$ and $S^{z}_{\text{tot}}=1/2$, we have five energy levels: $E_1(2,1/2)=E_2(2,1/2)=0$, $E_3(2,1/2)=(J_0-J_2)/2$, and \bea
E_{4,5}(2,1/2)&=&-\frac{J_0-J_2}2\nonumber\\
&&\pm\frac12\sqrt{3J_0(1-g^2)+(J_0-J_2)^2}.
\eea
The latter pair of eigenvalues becomes complex for $|g|>\sqrt{1+(1-J_2/J_0)^2/3}\geq1$. Therefore, for $|g|<1$ the entire spectrum of $H_I$ is real and the $\mc{PT}$ symmetry is preserved. For $|g|>1$, at least the eigenvalues in the $N_e=1$ sector  become complex and the   $\mc{PT}$ symmetry is spontaneously broken. 

Figure \ref{fig.4} illustrates the behavior of the energy levels $E_{l}(1,1)$ as a function of $g$. As we approach $g=1$ from below, the eigenvalues coalesce with the characteristic square-root dependence of exceptional points \cite{Heiss2012}.  For $|g|=1$,   the eigenvalues of $H_I$ depend only on the difference $J_0-J_2$. In particular, for $J_0=J_2$, all the eigenstates  become degenerate with  eigenvalue zero, implying that the impurity effectively decouples from the wires. Note that the condition $|g|=1$ and $J_0=J_2$  also corresponds to the special line (a separatrix) $\lambda_0=\lambda_1=\lambda_2$ in the weak-coupling RG flow shown in Fig.  \ref{fig.3}(c).

\section{Strong-coupling fixed point and conductance} \label{Transmittance}

When the effective Kondo couplings diverge in the low-energy limit, ${J_n(\Lambda)\to\infty}$ with $J_0(\Lambda)>J_2(\Lambda)$ along the flow, a conduction electron forms a singlet with the impurity spin.  The low-energy effective Hamiltonian for the remaining electrons in the wires  can be obtained by projecting out the $\mc{PT}$-symmetric orbital involved in the singlet state in   Eq. (\ref{singlet}). We introduce  the linear combinations \be
\tilde c_{\pm,\sigma}=\frac{1}{\sqrt2}\left(c_{-1,\sigma}\pm e^{-i\alpha} c_{1,\sigma}\right).
\ee
For $\alpha=0$, these are the annihilation operators for symmetric and antisymmetric orbitals (which are eigenstates of $\mc P$), respectively. In the more general $\mc{PT}$-symmetric problem, $c_{+,\sigma}$ annihilates an electron with spin $\sigma$ in the orbital state that becomes inaccessible at low energies where the singlet cannot be broken. We then define the projection operator $P$ onto the remaining electronic orbitals, such that $Pc_{-1}P=\tilde{c}_-/\sqrt2$ and $Pc_1P=-e^{i\alpha}\tilde{c}_-/\sqrt2$. The projection of the tight-binding Hamiltonian in the wires gives
\bea \label{eq.37}
H_{\text{sc}}&=&PH_0P\nonumber\\
&=&-t{\displaystyle\sum_{j\leq-3}\left(c_{j}^{\dagger}c^{\phantom\dagger}_{j+1}+\text{h.c.}\right)} -t{\displaystyle\sum_{j\geq2}\left(c_{j}^{\dagger}c^{\phantom\dagger}_{j+1}+\text{h.c.}\right)}\nonumber\\
&&-\frac{t}{\sqrt{2}}\left(c^{\dagger}_{-2}\tilde{c}^{\phantom\dagger}_{-}-e^{i\alpha}c^{\dagger}_{2}\tilde{c}_{-}+\text{h.c.}\right).
\eea
Note that the magnitude of hopping parameter is reduced by  a factor of $1/\sqrt2$ at the junction. Moreover, there is a phase factor $e^{i(\alpha+\pi)}$ associated with the (Hermitian) hopping process between the state annihilated by $\tilde c_{-}$ and the site $j=2$. This phase factor can be removed by performing the gauge transformation $c_{j}\to -e^{i\alpha}c_j $ for ${j\geq 2}$. We then obtain \bea
\tilde H_{\text{sc}}&=&-t{\displaystyle\sum_{j\leq-3}\left(c_{j}^{\dagger}c^{\phantom\dagger}_{j+1}+\text{h.c.}\right)} -t{\displaystyle\sum_{j\geq2}\left(c_{j}^{\dagger}c^{\phantom\dagger}_{j+1}+\text{h.c.}\right)}\nonumber\\
&&-\frac{t}{\sqrt{2}}\left(c^{\dagger}_{-2}\tilde{c}^{\phantom\dagger}_{-}+c^{\dagger}_{2}\tilde{c}_{-}+\text{h.c.}\right).\label{gaugetransf}
\eea
This is  now a $\mc{P}$- and $\mc{T}$-invariant  tight-binding model for a single infinite wire. Remarkably, it coincides with the effective Hamiltonian for the usual Kondo model in the strong coupling limit  \cite{Simon}. 

The linear conductance $G$ can be related to the transmission amplitude $ T$  through the junction using  the Landauer-B\"uttiker formalism \cite{Datta1997}:\be
G=\frac{2e^2}{h} T.
\ee 
At the strong coupling fixed point described by Hamiltonian (\ref{gaugetransf}), the transmission amplitude can be calculated by solving the single-particle scattering problem. Following Refs.  \cite{Simon,Pustilnik}, we obtain $ T =\sin^2(k_F)$ \cite{Simon}, which implies ideal transmittance $T=1$ in the particle-hole symmetric case $k_F=\pi/2$. This leads to the ideal conductance $G=2e^2/h$ at the strong-coupling  fixed point, as for  the usual Kondo effect in quantum dots \cite{Pustilnik2004}.

On the other hand, in the regime $|g|>1$, the effective  couplings $\lambda_n(\Lambda)$, with $n=0,1,2$,  vanish in the low-energy limit. In fact, it follows from the RG equations  (\ref{copyeq19}) and (\ref{eq23}) that they vanish logarithmically,  $\lambda_n(\Lambda)\sim [\ln(\Lambda_0/\Lambda)]^{-1}$ for $\Lambda\to0$, where $\Lambda_0\sim t$ is the bare cutoff scale. The conductance in this case can be calculated similarly to the weak coupling regime of the Kondo model, namely by starting from the Kubo formula and applying   second-order perturbation theory in the effective couplings (see Refs. \cite{Pustilnik2004,Tai} for details).  Therefore, in the $\mc{PT}$-broken regime $|g|>1$, the conductance scales as $G(\Lambda)\sim  [\ln(\Lambda_0/\Lambda)]^{-2}$ and vanishes at the local-moment fixed point,  at which the wires decouple from the impurity and the currents are totally reflected at the boundary.

Within the effective field theory, the strong-coupling fixed point can be understood as a $\pi/2$ phase shift that changes the boundary conditions of electron states in the channel involved in the Kondo coupling \cite{Affleck:1995}. The effective field theory can also be used to show that the Kondo fixed point is stable. Since the impurity spin disappears from the low-energy effective Hamiltonian, the  perturbations to the Kondo fixed point are all irrelevant boundary operators and the low-energy properties are described by a  local Fermi liquid theory \cite{Nozieres1974}. No relevant perturbations arise  in the non-Hermitian model with  $\mc{PT}$, SU(2) and particle-hole symmetries. Particle-hole symmetry breaking allows for marginal perturbations that reduce the conductance from the  ideal to a lower  non-universal value. In the Hermitian Kondo model, this marginal perturbation corresponds to  the $s$-wave potential scattering term $V_0\Psi^\dagger(0)\Psi(0)$ \cite{hewson_1993}. In the non-Hermitian model without particle-hole symmetry, we have an additional marginal perturbation allowed by $\mc{PT}$ symmetry, represented  by $V_1\Psi^\dagger(0)\Upsilon\Psi(0)$ (where $\Upsilon$ is the imaginary antisymmetric matrix in Eq. (\ref{eq17})).  


Finally, we comment on the possibility of varying the parameter $g$ across the exceptional point $g=1$. Using the perturbative expressions for the bare  exchange couplings in Eqs. (\ref{Jcoupling1})-(\ref{Jcoupling3}), we can show that $g=\lambda_1/\lambda_0= J_1/J_0=\sin(2\theta)$ with \be
\theta=\arctan\left[\frac{\sqrt{JJ'}\sin\phi}{J+\sqrt{JJ'}\cos\phi}\right].\ee
Therefore, these perturbative expressions predict $|g| \leq 1$. However, this relation does not hold  beyond second-order perturbation theory in $t'$ and $w$ or for a more general lattice model than Eq. (\ref{eq1}) (for instance, including non-Hermitian hopping processes  between the dot and the second site in each wire).  More generally, the  bare coupling constants $\lambda_n$ that set the initial values in the RG flow must be treated as independent phenomenological parameters. Nonetheless, the above result suggests that the  spontaneous breaking of  $\mc{PT}$ symmetry in our non-Hermitian Kondo model should be difficult to realize in the regime  $t',w\ll |\epsilon_d|,U$. Instead, one should look for stronger tunnelling  between the wires and the quantum dot, but still in the Coulomb blockade regime where charge fluctuations in the dot can be neglected.

\section{Conclusions}\label{Conclusions}
We have investigated an  Anderson  impurity model  with  $\mathcal{PT}$-symmetric  non-Hermitian hopping between the wires and the localized state in the quantum dot. Using a Schrieffer-Wolff transformation, we obtained the $\mc{PT}$-symmetric Kondo model that describes the coupling to the impurity spin. Our perturbative renormalization group analysis showed that the fate of the Kondo effect is controlled by the parameter $g$ defined as  the ratio between the non-Hermitian   coupling and the usual single-channel Kondo coupling. For $|g|<1$, the spectrum of the Kondo interaction is real and the Kondo effect persists. In the particle-hole symmetric case, the strong coupling fixed point of the $\mc{PT}$-symmetric Kondo model has ideal conductance through the quantum dot. For $|g|>1$, the spectrum becomes complex and the $\mathcal{PT}$ symmetry  is spontaneously broken. In this case, the low-energy physics is governed by a local-moment fixed point with zero conductance. 

Some   open questions include the generalization to the multichannel Kondo model \cite{Nozieresph,AffleckLudwig} and the interplay of $\mc{PT}$-symmetric interactions at the boundary and  in the bulk, as in the Kondo effect in Tomonaga-Luttinger liquids \cite{Lee1992,Furusaki1994}. More generally, it would be  interesting to search for new boundary  fixed points  unique to $\mc{PT}$-symmetric non-Hermitian systems, perhaps with  chiral transport properties analogous to those realized   in quantum optics \cite{Lodahl,Longhireview}. To go beyond  perturbative approaches, it would be interesting to generalize powerful numerical techniques that have been instrumental  in the study of   quantum impurity models, such as Wilson's numerical  renormalization group \cite{Wilson1975,Bulla2008}.

\begin{acknowledgements}
We thank A. Ferraz, T. Macri and L. G. G. V. Dias da Silva  for helpful discussions. We acknowledge financial support from CNPq and the Brazilian ministries MEC and MCTIC.
\end{acknowledgements}
 
\bibliographystyle{apsrev4-1}
\bibliography{Reference}

\begin{thebibliography}{56}%
\makeatletter
\providecommand \@ifxundefined [1]{%
 \@ifx{#1\undefined}
}%
\providecommand \@ifnum [1]{%
 \ifnum #1\expandafter \@firstoftwo
 \else \expandafter \@secondoftwo
 \fi
}%
\providecommand \@ifx [1]{%
 \ifx #1\expandafter \@firstoftwo
 \else \expandafter \@secondoftwo
 \fi
}%
\providecommand \natexlab [1]{#1}%
\providecommand \enquote  [1]{``#1''}%
\providecommand \bibnamefont  [1]{#1}%
\providecommand \bibfnamefont [1]{#1}%
\providecommand \citenamefont [1]{#1}%
\providecommand \href@noop [0]{\@secondoftwo}%
\providecommand \href [0]{\begingroup \@sanitize@url \@href}%
\providecommand \@href[1]{\@@startlink{#1}\@@href}%
\providecommand \@@href[1]{\endgroup#1\@@endlink}%
\providecommand \@sanitize@url [0]{\catcode `\\12\catcode `\$12\catcode
  `\&12\catcode `\#12\catcode `\^12\catcode `\_12\catcode `\%12\relax}%
\providecommand \@@startlink[1]{}%
\providecommand \@@endlink[0]{}%
\providecommand \url  [0]{\begingroup\@sanitize@url \@url }%
\providecommand \@url [1]{\endgroup\@href {#1}{\urlprefix }}%
\providecommand \urlprefix  [0]{URL }%
\providecommand \Eprint [0]{\href }%
\providecommand \doibase [0]{http://dx.doi.org/}%
\providecommand \selectlanguage [0]{\@gobble}%
\providecommand \bibinfo  [0]{\@secondoftwo}%
\providecommand \bibfield  [0]{\@secondoftwo}%
\providecommand \translation [1]{[#1]}%
\providecommand \BibitemOpen [0]{}%
\providecommand \bibitemStop [0]{}%
\providecommand \bibitemNoStop [0]{.\EOS\space}%
\providecommand \EOS [0]{\spacefactor3000\relax}%
\providecommand \BibitemShut  [1]{\csname bibitem#1\endcsname}%
\let\auto@bib@innerbib\@empty
\bibitem [{\citenamefont {Moiseyev}(2011)}]{moiseyev2011non}%
  \BibitemOpen
  \bibfield  {author} {\bibinfo {author} {\bibfnamefont {N.}~\bibnamefont
  {Moiseyev}},\ }\href@noop {} {\emph {\bibinfo {title} {Non-Hermitian Quantum
  Mechanics}}}\ (\bibinfo  {publisher} {Cambridge University Press},\ \bibinfo
  {year} {2011})\BibitemShut {NoStop}%
\bibitem [{\citenamefont {Rotter}(2009)}]{Rotter}%
  \BibitemOpen
  \bibfield  {author} {\bibinfo {author} {\bibfnamefont {I.}~\bibnamefont
  {Rotter}},\ }\href {http://stacks.iop.org/1751-8121/42/i=15/a=153001}
  {\bibfield  {journal} {\bibinfo  {journal} {J. Phys. A: Math. and Theor.}\
  }\textbf {\bibinfo {volume} {42}},\ \bibinfo {pages} {153001} (\bibinfo
  {year} {2009})}\BibitemShut {NoStop}%
\bibitem [{\citenamefont {Plenio}\ and\ \citenamefont
  {Knight}(1998)}]{Plenio1998}%
  \BibitemOpen
  \bibfield  {author} {\bibinfo {author} {\bibfnamefont {M.~B.}\ \bibnamefont
  {Plenio}}\ and\ \bibinfo {author} {\bibfnamefont {P.~L.}\ \bibnamefont
  {Knight}},\ }\href {\doibase 10.1103/RevModPhys.70.101} {\bibfield  {journal}
  {\bibinfo  {journal} {Rev. Mod. Phys.}\ }\textbf {\bibinfo {volume} {70}},\
  \bibinfo {pages} {101} (\bibinfo {year} {1998})}\BibitemShut {NoStop}%
\bibitem [{\citenamefont {Hatano}\ and\ \citenamefont
  {Nelson}(1997)}]{Hatano1997}%
  \BibitemOpen
  \bibfield  {author} {\bibinfo {author} {\bibfnamefont {N.}~\bibnamefont
  {Hatano}}\ and\ \bibinfo {author} {\bibfnamefont {D.~R.}\ \bibnamefont
  {Nelson}},\ }\href {\doibase 10.1103/PhysRevB.56.8651} {\bibfield  {journal}
  {\bibinfo  {journal} {Phys. Rev. B}\ }\textbf {\bibinfo {volume} {56}},\
  \bibinfo {pages} {8651} (\bibinfo {year} {1997})}\BibitemShut {NoStop}%
\bibitem [{\citenamefont {Bender}\ and\ \citenamefont
  {Boettcher}(1998)}]{Bender}%
  \BibitemOpen
  \bibfield  {author} {\bibinfo {author} {\bibfnamefont {C.~M.}\ \bibnamefont
  {Bender}}\ and\ \bibinfo {author} {\bibfnamefont {S.}~\bibnamefont
  {Boettcher}},\ }\href {\doibase 10.1103/PhysRevLett.80.5243} {\bibfield
  {journal} {\bibinfo  {journal} {Phys. Rev. Lett.}\ }\textbf {\bibinfo
  {volume} {80}},\ \bibinfo {pages} {5243} (\bibinfo {year}
  {1998})}\BibitemShut {NoStop}%
\bibitem [{\citenamefont {Heiss}(2012)}]{Heiss2012}%
  \BibitemOpen
  \bibfield  {author} {\bibinfo {author} {\bibfnamefont {W.~D.}\ \bibnamefont
  {Heiss}},\ }\href {http://stacks.iop.org/1751-8121/45/i=44/a=444016}
  {\bibfield  {journal} {\bibinfo  {journal} {J. Phys. A: Math. and Theor.}\
  }\textbf {\bibinfo {volume} {45}},\ \bibinfo {pages} {444016} (\bibinfo
  {year} {2012})}\BibitemShut {NoStop}%
\bibitem [{\citenamefont {Bender}\ and\ \citenamefont {Darg}(2007)}]{Bender2}%
  \BibitemOpen
  \bibfield  {author} {\bibinfo {author} {\bibfnamefont {C.~M.}\ \bibnamefont
  {Bender}}\ and\ \bibinfo {author} {\bibfnamefont {D.~W.}\ \bibnamefont
  {Darg}},\ }\href {\doibase 10.1063/1.2720279} {\bibfield  {journal} {\bibinfo
   {journal} {J. Math. Phys.}\ }\textbf {\bibinfo {volume} {48}},\ \bibinfo
  {pages} {042703} (\bibinfo {year} {2007})}\BibitemShut {NoStop}%
\bibitem [{\citenamefont {Feng}\ \emph {et~al.}(2017)\citenamefont {Feng},
  \citenamefont {El-Ganainy},\ and\ \citenamefont {Ge}}]{Feng2017}%
  \BibitemOpen
  \bibfield  {author} {\bibinfo {author} {\bibfnamefont {L.}~\bibnamefont
  {Feng}}, \bibinfo {author} {\bibfnamefont {R.}~\bibnamefont {El-Ganainy}}, \
  and\ \bibinfo {author} {\bibfnamefont {L.}~\bibnamefont {Ge}},\ }\href
  {\doibase 10.1038/s41566-017-0031-1} {\bibfield  {journal} {\bibinfo
  {journal} {Nat. Phot.}\ }\textbf {\bibinfo {volume} {11}},\ \bibinfo {pages}
  {752} (\bibinfo {year} {2017})}\BibitemShut {NoStop}%
\bibitem [{\citenamefont {El-Ganainy}\ \emph {et~al.}(2018)\citenamefont
  {El-Ganainy}, \citenamefont {Makris}, \citenamefont {Khajavikhan},
  \citenamefont {Musslimani}, \citenamefont {Rotter},\ and\ \citenamefont
  {Christodoulides}}]{Ramy2018}%
  \BibitemOpen
  \bibfield  {author} {\bibinfo {author} {\bibfnamefont {R.}~\bibnamefont
  {El-Ganainy}}, \bibinfo {author} {\bibfnamefont {K.~G.}\ \bibnamefont
  {Makris}}, \bibinfo {author} {\bibfnamefont {M.}~\bibnamefont {Khajavikhan}},
  \bibinfo {author} {\bibfnamefont {Z.~H.}\ \bibnamefont {Musslimani}},
  \bibinfo {author} {\bibfnamefont {S.}~\bibnamefont {Rotter}}, \ and\ \bibinfo
  {author} {\bibfnamefont {D.~N.}\ \bibnamefont {Christodoulides}},\ }\href
  {\doibase 10.1038/nphys4323} {\bibfield  {journal} {\bibinfo  {journal} {Nat.
  Phys.}\ }\textbf {\bibinfo {volume} {14}},\ \bibinfo {pages} {11} (\bibinfo
  {year} {2018})}\BibitemShut {NoStop}%
\bibitem [{\citenamefont {Longhi}(2017)}]{Longhireview}%
  \BibitemOpen
  \bibfield  {author} {\bibinfo {author} {\bibfnamefont {S.}~\bibnamefont
  {Longhi}},\ }\href {http://stacks.iop.org/0295-5075/120/i=6/a=64001}
  {\bibfield  {journal} {\bibinfo  {journal} {EPL (Europhysics Letters)}\
  }\textbf {\bibinfo {volume} {120}},\ \bibinfo {pages} {64001} (\bibinfo
  {year} {2017})}\BibitemShut {NoStop}%
\bibitem [{\citenamefont {Guo}\ \emph {et~al.}(2009)\citenamefont {Guo},
  \citenamefont {Salamo}, \citenamefont {Duchesne}, \citenamefont {Morandotti},
  \citenamefont {Volatier-Ravat}, \citenamefont {Aimez}, \citenamefont
  {Siviloglou},\ and\ \citenamefont {Christodoulides}}]{Guo}%
  \BibitemOpen
  \bibfield  {author} {\bibinfo {author} {\bibfnamefont {A.}~\bibnamefont
  {Guo}}, \bibinfo {author} {\bibfnamefont {G.~J.}\ \bibnamefont {Salamo}},
  \bibinfo {author} {\bibfnamefont {D.}~\bibnamefont {Duchesne}}, \bibinfo
  {author} {\bibfnamefont {R.}~\bibnamefont {Morandotti}}, \bibinfo {author}
  {\bibfnamefont {M.}~\bibnamefont {Volatier-Ravat}}, \bibinfo {author}
  {\bibfnamefont {V.}~\bibnamefont {Aimez}}, \bibinfo {author} {\bibfnamefont
  {G.~A.}\ \bibnamefont {Siviloglou}}, \ and\ \bibinfo {author} {\bibfnamefont
  {D.~N.}\ \bibnamefont {Christodoulides}},\ }\href
  {https://link.aps.org/doi/10.1103/PhysRevLett.103.093902} {\bibfield
  {journal} {\bibinfo  {journal} {Phys. Rev. Lett.}\ }\textbf {\bibinfo
  {volume} {103}},\ \bibinfo {pages} {093902} (\bibinfo {year}
  {2009})}\BibitemShut {NoStop}%
\bibitem [{\citenamefont {R\"{u}ter}\ \emph {et~al.}(2010)\citenamefont
  {R\"{u}ter}, \citenamefont {Makris}, \citenamefont {El-Ganainy},
  \citenamefont {Christodoulides}, \citenamefont {Segev},\ and\ \citenamefont
  {Kip}}]{Ruter}%
  \BibitemOpen
  \bibfield  {author} {\bibinfo {author} {\bibfnamefont {C.~E.}\ \bibnamefont
  {R\"{u}ter}}, \bibinfo {author} {\bibfnamefont {K.~G.}\ \bibnamefont
  {Makris}}, \bibinfo {author} {\bibfnamefont {R.~R.}\ \bibnamefont
  {El-Ganainy}}, \bibinfo {author} {\bibfnamefont {D.~N.}\ \bibnamefont
  {Christodoulides}}, \bibinfo {author} {\bibfnamefont {M.}~\bibnamefont
  {Segev}}, \ and\ \bibinfo {author} {\bibfnamefont {D.}~\bibnamefont {Kip}},\
  }\href {http://dx.doi.org/10.1038/nphys1515} {\bibfield  {journal} {\bibinfo
  {journal} {Nat. Phys.}\ }\textbf {\bibinfo {volume} {6}},\ \bibinfo {pages}
  {192} (\bibinfo {year} {2010})}\BibitemShut {NoStop}%
\bibitem [{\citenamefont {Zhang}\ \emph {et~al.}(2016)\citenamefont {Zhang},
  \citenamefont {Zhang}, \citenamefont {Sheng}, \citenamefont {Yang},
  \citenamefont {Miri}, \citenamefont {Christodoulides}, \citenamefont {He},
  \citenamefont {Zhang},\ and\ \citenamefont {Xiao}}]{Zhang}%
  \BibitemOpen
  \bibfield  {author} {\bibinfo {author} {\bibfnamefont {Z.}~\bibnamefont
  {Zhang}}, \bibinfo {author} {\bibfnamefont {Y.}~\bibnamefont {Zhang}},
  \bibinfo {author} {\bibfnamefont {J.}~\bibnamefont {Sheng}}, \bibinfo
  {author} {\bibfnamefont {L.}~\bibnamefont {Yang}}, \bibinfo {author}
  {\bibfnamefont {M.-A.}\ \bibnamefont {Miri}}, \bibinfo {author}
  {\bibfnamefont {D.~N.}\ \bibnamefont {Christodoulides}}, \bibinfo {author}
  {\bibfnamefont {B.}~\bibnamefont {He}}, \bibinfo {author} {\bibfnamefont
  {Y.}~\bibnamefont {Zhang}}, \ and\ \bibinfo {author} {\bibfnamefont
  {M.}~\bibnamefont {Xiao}},\ }\href {\doibase 10.1103/PhysRevLett.117.123601}
  {\bibfield  {journal} {\bibinfo  {journal} {Phys. Rev. Lett.}\ }\textbf
  {\bibinfo {volume} {117}},\ \bibinfo {pages} {123601} (\bibinfo {year}
  {2016})}\BibitemShut {NoStop}%
\bibitem [{\citenamefont {Peng}\ \emph {et~al.}(2014)\citenamefont {Peng},
  \citenamefont {{\"O}zdemir}, \citenamefont {Lei}, \citenamefont {Monifi},
  \citenamefont {Gianfreda}, \citenamefont {Long}, \citenamefont {Fan},
  \citenamefont {Nori}, \citenamefont {Bender},\ and\ \citenamefont
  {Yang}}]{Peng2014}%
  \BibitemOpen
  \bibfield  {author} {\bibinfo {author} {\bibfnamefont {B.}~\bibnamefont
  {Peng}}, \bibinfo {author} {\bibfnamefont {{\c S}.~K.}\ \bibnamefont
  {{\"O}zdemir}}, \bibinfo {author} {\bibfnamefont {F.}~\bibnamefont {Lei}},
  \bibinfo {author} {\bibfnamefont {F.}~\bibnamefont {Monifi}}, \bibinfo
  {author} {\bibfnamefont {M.}~\bibnamefont {Gianfreda}}, \bibinfo {author}
  {\bibfnamefont {G.~L.}\ \bibnamefont {Long}}, \bibinfo {author}
  {\bibfnamefont {S.}~\bibnamefont {Fan}}, \bibinfo {author} {\bibfnamefont
  {F.}~\bibnamefont {Nori}}, \bibinfo {author} {\bibfnamefont {C.~M.}\
  \bibnamefont {Bender}}, \ and\ \bibinfo {author} {\bibfnamefont
  {L.}~\bibnamefont {Yang}},\ }\href {http://dx.doi.org/10.1038/nphys2927}
  {\bibfield  {journal} {\bibinfo  {journal} {Nat. Phys.}\ }\textbf {\bibinfo
  {volume} {10}},\ \bibinfo {pages} {394} (\bibinfo {year} {2014})}\BibitemShut
  {NoStop}%
\bibitem [{\citenamefont {Shi}\ \emph {et~al.}(2016)\citenamefont {Shi},
  \citenamefont {Dubois}, \citenamefont {Chen}, \citenamefont {Cheng},
  \citenamefont {Ramezani}, \citenamefont {Wang},\ and\ \citenamefont
  {Zhang}}]{Shi2016}%
  \BibitemOpen
  \bibfield  {author} {\bibinfo {author} {\bibfnamefont {C.}~\bibnamefont
  {Shi}}, \bibinfo {author} {\bibfnamefont {M.}~\bibnamefont {Dubois}},
  \bibinfo {author} {\bibfnamefont {Y.}~\bibnamefont {Chen}}, \bibinfo {author}
  {\bibfnamefont {L.}~\bibnamefont {Cheng}}, \bibinfo {author} {\bibfnamefont
  {H.}~\bibnamefont {Ramezani}}, \bibinfo {author} {\bibfnamefont
  {Y.}~\bibnamefont {Wang}}, \ and\ \bibinfo {author} {\bibfnamefont
  {X.}~\bibnamefont {Zhang}},\ }\href {http://dx.doi.org/10.1038/ncomms11110}
  {\bibfield  {journal} {\bibinfo  {journal} {Nat. Comm.}\ }\textbf {\bibinfo
  {volume} {7}},\ \bibinfo {pages} {11110 EP } (\bibinfo {year}
  {2016})}\BibitemShut {NoStop}%
\bibitem [{\citenamefont {Aur\'egan}\ and\ \citenamefont
  {Pagneux}(2017)}]{Auregan2017}%
  \BibitemOpen
  \bibfield  {author} {\bibinfo {author} {\bibfnamefont {Y.}~\bibnamefont
  {Aur\'egan}}\ and\ \bibinfo {author} {\bibfnamefont {V.}~\bibnamefont
  {Pagneux}},\ }\href {\doibase 10.1103/PhysRevLett.118.174301} {\bibfield
  {journal} {\bibinfo  {journal} {Phys. Rev. Lett.}\ }\textbf {\bibinfo
  {volume} {118}},\ \bibinfo {pages} {174301} (\bibinfo {year}
  {2017})}\BibitemShut {NoStop}%
\bibitem [{\citenamefont {Quijandr\'{\i}a}\ \emph {et~al.}(2018)\citenamefont
  {Quijandr\'{\i}a}, \citenamefont {Naether}, \citenamefont {\"Ozdemir},
  \citenamefont {Nori},\ and\ \citenamefont {Zueco}}]{Quijandria2018}%
  \BibitemOpen
  \bibfield  {author} {\bibinfo {author} {\bibfnamefont {F.}~\bibnamefont
  {Quijandr\'{\i}a}}, \bibinfo {author} {\bibfnamefont {U.}~\bibnamefont
  {Naether}}, \bibinfo {author} {\bibfnamefont {S.~K.}\ \bibnamefont
  {\"Ozdemir}}, \bibinfo {author} {\bibfnamefont {F.}~\bibnamefont {Nori}}, \
  and\ \bibinfo {author} {\bibfnamefont {D.}~\bibnamefont {Zueco}},\ }\href
  {https://link.aps.org/doi/10.1103/PhysRevA.97.053846} {\bibfield  {journal}
  {\bibinfo  {journal} {Phys. Rev. A}\ }\textbf {\bibinfo {volume} {97}},\
  \bibinfo {pages} {053846} (\bibinfo {year} {2018})}\BibitemShut {NoStop}%
\bibitem [{\citenamefont {Korff}\ and\ \citenamefont
  {Weston}(2007)}]{Korff2007}%
  \BibitemOpen
  \bibfield  {author} {\bibinfo {author} {\bibfnamefont {C.}~\bibnamefont
  {Korff}}\ and\ \bibinfo {author} {\bibfnamefont {R.}~\bibnamefont {Weston}},\
  }\href {http://stacks.iop.org/1751-8121/40/i=30/a=016} {\bibfield  {journal}
  {\bibinfo  {journal} {J. Phys. A: Math. and Theor.}\ }\textbf {\bibinfo
  {volume} {40}},\ \bibinfo {pages} {8845} (\bibinfo {year}
  {2007})}\BibitemShut {NoStop}%
\bibitem [{\citenamefont {Castro-Alvaredo}\ and\ \citenamefont
  {Fring}(2009)}]{Olalla2009}%
  \BibitemOpen
  \bibfield  {author} {\bibinfo {author} {\bibfnamefont {O.~A.}\ \bibnamefont
  {Castro-Alvaredo}}\ and\ \bibinfo {author} {\bibfnamefont {A.}~\bibnamefont
  {Fring}},\ }\href {http://stacks.iop.org/1751-8121/42/i=46/a=465211}
  {\bibfield  {journal} {\bibinfo  {journal} {J. Phys. A: Math. and Theor.}\
  }\textbf {\bibinfo {volume} {42}},\ \bibinfo {pages} {465211} (\bibinfo
  {year} {2009})}\BibitemShut {NoStop}%
\bibitem [{\citenamefont {Deguchi}\ and\ \citenamefont
  {Ghosh}(2009)}]{Deguchi2009}%
  \BibitemOpen
  \bibfield  {author} {\bibinfo {author} {\bibfnamefont {T.}~\bibnamefont
  {Deguchi}}\ and\ \bibinfo {author} {\bibfnamefont {P.~K.}\ \bibnamefont
  {Ghosh}},\ }\href {http://stacks.iop.org/1751-8121/42/i=47/a=475208}
  {\bibfield  {journal} {\bibinfo  {journal} {J. Phys. A: Math. and Theor.}\
  }\textbf {\bibinfo {volume} {42}},\ \bibinfo {pages} {475208} (\bibinfo
  {year} {2009})}\BibitemShut {NoStop}%
\bibitem [{\citenamefont {Shen}\ \emph {et~al.}(2018)\citenamefont {Shen},
  \citenamefont {Zhen},\ and\ \citenamefont {Fu}}]{Shen2018}%
  \BibitemOpen
  \bibfield  {author} {\bibinfo {author} {\bibfnamefont {H.}~\bibnamefont
  {Shen}}, \bibinfo {author} {\bibfnamefont {B.}~\bibnamefont {Zhen}}, \ and\
  \bibinfo {author} {\bibfnamefont {L.}~\bibnamefont {Fu}},\ }\href
  {https://link.aps.org/doi/10.1103/PhysRevLett.120.146402} {\bibfield
  {journal} {\bibinfo  {journal} {Phys. Rev. Lett.}\ }\textbf {\bibinfo
  {volume} {120}},\ \bibinfo {pages} {146402} (\bibinfo {year}
  {2018})}\BibitemShut {NoStop}%
\bibitem [{\citenamefont {Kawabata}\ \emph {et~al.}(2018)\citenamefont
  {Kawabata}, \citenamefont {Ashida}, \citenamefont {Katsura},\ and\
  \citenamefont {Ueda}}]{Kawabata}%
  \BibitemOpen
  \bibfield  {author} {\bibinfo {author} {\bibfnamefont {K.}~\bibnamefont
  {Kawabata}}, \bibinfo {author} {\bibfnamefont {Y.}~\bibnamefont {Ashida}},
  \bibinfo {author} {\bibfnamefont {H.}~\bibnamefont {Katsura}}, \ and\
  \bibinfo {author} {\bibfnamefont {M.}~\bibnamefont {Ueda}},\ }\href
  {https://link.aps.org/doi/10.1103/PhysRevB.98.085116} {\bibfield  {journal}
  {\bibinfo  {journal} {Phys. Rev. B}\ }\textbf {\bibinfo {volume} {98}},\
  \bibinfo {pages} {085116} (\bibinfo {year} {2018})}\BibitemShut {NoStop}%
\bibitem [{\citenamefont {{Yao}}\ and\ \citenamefont {{Wang}}(2018)}]{Yao2018}%
  \BibitemOpen
  \bibfield  {author} {\bibinfo {author} {\bibfnamefont {S.}~\bibnamefont
  {{Yao}}}\ and\ \bibinfo {author} {\bibfnamefont {Z.}~\bibnamefont {{Wang}}},\
  }\href@noop {} {\bibfield  {journal} {\bibinfo  {journal} {ArXiv e-prints}\ }
  (\bibinfo {year} {2018})},\ \Eprint {http://arxiv.org/abs/1803.01876}
  {arXiv:1803.01876 [cond-mat.mes-hall]} \BibitemShut {NoStop}%
\bibitem [{\citenamefont {Li}\ \emph {et~al.}(2018)\citenamefont {Li},
  \citenamefont {Zhang}, \citenamefont {Zhang},\ and\ \citenamefont
  {Song}}]{Li2018}%
  \BibitemOpen
  \bibfield  {author} {\bibinfo {author} {\bibfnamefont {C.}~\bibnamefont
  {Li}}, \bibinfo {author} {\bibfnamefont {X.~Z.}\ \bibnamefont {Zhang}},
  \bibinfo {author} {\bibfnamefont {G.}~\bibnamefont {Zhang}}, \ and\ \bibinfo
  {author} {\bibfnamefont {Z.}~\bibnamefont {Song}},\ }\href
  {https://link.aps.org/doi/10.1103/PhysRevB.97.115436} {\bibfield  {journal}
  {\bibinfo  {journal} {Phys. Rev. B}\ }\textbf {\bibinfo {volume} {97}},\
  \bibinfo {pages} {115436} (\bibinfo {year} {2018})}\BibitemShut {NoStop}%
\bibitem [{\citenamefont {{Gong}}\ \emph {et~al.}(2018)\citenamefont {{Gong}},
  \citenamefont {{Ashida}}, \citenamefont {{Kawabata}}, \citenamefont
  {{Takasan}}, \citenamefont {{Higashikawa}},\ and\ \citenamefont
  {{Ueda}}}]{Gong2018}%
  \BibitemOpen
  \bibfield  {author} {\bibinfo {author} {\bibfnamefont {Z.}~\bibnamefont
  {{Gong}}}, \bibinfo {author} {\bibfnamefont {Y.}~\bibnamefont {{Ashida}}},
  \bibinfo {author} {\bibfnamefont {K.}~\bibnamefont {{Kawabata}}}, \bibinfo
  {author} {\bibfnamefont {K.}~\bibnamefont {{Takasan}}}, \bibinfo {author}
  {\bibfnamefont {S.}~\bibnamefont {{Higashikawa}}}, \ and\ \bibinfo {author}
  {\bibfnamefont {M.}~\bibnamefont {{Ueda}}},\ }\href@noop {} {\bibfield
  {journal} {\bibinfo  {journal} {ArXiv e-prints}\ } (\bibinfo {year}
  {2018})},\ \Eprint {http://arxiv.org/abs/1802.07964} {arXiv:1802.07964
  [cond-mat.mes-hall]} \BibitemShut {NoStop}%
\bibitem [{\citenamefont {Bender}\ \emph {et~al.}(2004)\citenamefont {Bender},
  \citenamefont {Brody},\ and\ \citenamefont {Jones}}]{BenderPRD}%
  \BibitemOpen
  \bibfield  {author} {\bibinfo {author} {\bibfnamefont {C.~M.}\ \bibnamefont
  {Bender}}, \bibinfo {author} {\bibfnamefont {D.~C.}\ \bibnamefont {Brody}}, \
  and\ \bibinfo {author} {\bibfnamefont {H.~F.}\ \bibnamefont {Jones}},\ }\href
  {\doibase 10.1103/PhysRevD.70.025001} {\bibfield  {journal} {\bibinfo
  {journal} {Phys. Rev. D}\ }\textbf {\bibinfo {volume} {70}},\ \bibinfo
  {pages} {025001} (\bibinfo {year} {2004})}\BibitemShut {NoStop}%
\bibitem [{\citenamefont {Ashida}\ \emph {et~al.}(2017)\citenamefont {Ashida},
  \citenamefont {Furukawa},\ and\ \citenamefont {Ueda}}]{Ashida}%
  \BibitemOpen
  \bibfield  {author} {\bibinfo {author} {\bibfnamefont {Y.}~\bibnamefont
  {Ashida}}, \bibinfo {author} {\bibfnamefont {S.}~\bibnamefont {Furukawa}}, \
  and\ \bibinfo {author} {\bibfnamefont {M.}~\bibnamefont {Ueda}},\ }\href
  {\doibase 10.1038/ncomms15791} {\bibfield  {journal} {\bibinfo  {journal}
  {Nat. Comm.}\ }\textbf {\bibinfo {volume} {8}},\ \bibinfo {pages} {15791}
  (\bibinfo {year} {2017})}\BibitemShut {NoStop}%
\bibitem [{\citenamefont {Affleck}(2008)}]{Affleck2008}%
  \BibitemOpen
  \bibfield  {author} {\bibinfo {author} {\bibfnamefont {I.}~\bibnamefont
  {Affleck}},\ }\enquote {\bibinfo {title} {Quantum impurity problems in
  condensed matter physics},}\ in\ \href@noop {} {\emph {\bibinfo {booktitle}
  {Exact Methods in Low-Dimensional Statistical Physics and Quantum
  Computing}}},\ \bibinfo {series and number} {Proceedings of the Les Houches
  Summer School 2008, Session LXXXIX},\ \bibinfo {editor} {edited by\ \bibinfo
  {editor} {\bibfnamefont {J.}~\bibnamefont {Jacobsen}}, \bibinfo {editor}
  {\bibfnamefont {S.}~\bibnamefont {Ouvry}}, \ and\ \bibinfo {editor}
  {\bibfnamefont {V.}~\bibnamefont {Pasquier}}}\ (\bibinfo  {publisher} {Oxford
  University Press},\ \bibinfo {year} {2008})\BibitemShut {NoStop}%
\bibitem [{\citenamefont {Anderson}(1961)}]{Anderson}%
  \BibitemOpen
  \bibfield  {author} {\bibinfo {author} {\bibfnamefont {P.~W.}\ \bibnamefont
  {Anderson}},\ }\href {\doibase 10.1103/PhysRev.124.41} {\bibfield  {journal}
  {\bibinfo  {journal} {Phys. Rev.}\ }\textbf {\bibinfo {volume} {124}},\
  \bibinfo {pages} {41} (\bibinfo {year} {1961})}\BibitemShut {NoStop}%
\bibitem [{\citenamefont {Pustilnik}\ and\ \citenamefont
  {Glazman}(2004)}]{Pustilnik2004}%
  \BibitemOpen
  \bibfield  {author} {\bibinfo {author} {\bibfnamefont {M.}~\bibnamefont
  {Pustilnik}}\ and\ \bibinfo {author} {\bibfnamefont {L.}~\bibnamefont
  {Glazman}},\ }\href {http://stacks.iop.org/0953-8984/16/i=16/a=R01}
  {\bibfield  {journal} {\bibinfo  {journal} {J. Phys.: Condens. Matter}\
  }\textbf {\bibinfo {volume} {16}},\ \bibinfo {pages} {R513} (\bibinfo {year}
  {2004})}\BibitemShut {NoStop}%
\bibitem [{\citenamefont {Kondo}(1964)}]{Kondo}%
  \BibitemOpen
  \bibfield  {author} {\bibinfo {author} {\bibfnamefont {J.}~\bibnamefont
  {Kondo}},\ }\href {\doibase 10.1143/PTP.32.37} {\bibfield  {journal}
  {\bibinfo  {journal} {Prog. Theor. Phys.}\ }\textbf {\bibinfo {volume}
  {32}},\ \bibinfo {pages} {37} (\bibinfo {year} {1964})}\BibitemShut {NoStop}%
\bibitem [{\citenamefont {Hewson}(1993)}]{hewson_1993}%
  \BibitemOpen
  \bibfield  {author} {\bibinfo {author} {\bibfnamefont {A.~C.}\ \bibnamefont
  {Hewson}},\ }\href {\doibase 10.1017/CBO9780511470752} {\emph {\bibinfo
  {title} {The Kondo Problem to Heavy Fermions}}},\ Cambridge Studies in
  Magnetism\ (\bibinfo  {publisher} {Cambridge University Press},\ \bibinfo
  {year} {1993})\BibitemShut {NoStop}%
\bibitem [{\citenamefont {Goldhaber-Gordon}\ \emph {et~al.}(1998)\citenamefont
  {Goldhaber-Gordon}, \citenamefont {Shtrikman}, \citenamefont {Mahalu},
  \citenamefont {Abusch-Magder}, \citenamefont {Meirav},\ and\ \citenamefont
  {Kastner}}]{GoldhaberGordon1998}%
  \BibitemOpen
  \bibfield  {author} {\bibinfo {author} {\bibfnamefont {D.}~\bibnamefont
  {Goldhaber-Gordon}}, \bibinfo {author} {\bibfnamefont {H.}~\bibnamefont
  {Shtrikman}}, \bibinfo {author} {\bibfnamefont {D.}~\bibnamefont {Mahalu}},
  \bibinfo {author} {\bibfnamefont {D.}~\bibnamefont {Abusch-Magder}}, \bibinfo
  {author} {\bibfnamefont {U.}~\bibnamefont {Meirav}}, \ and\ \bibinfo {author}
  {\bibfnamefont {M.~A.}\ \bibnamefont {Kastner}},\ }\href
  {http://dx.doi.org/10.1038/34373} {\bibfield  {journal} {\bibinfo  {journal}
  {Nature}\ }\textbf {\bibinfo {volume} {391}},\ \bibinfo {pages} {156}
  (\bibinfo {year} {1998})}\BibitemShut {NoStop}%
\bibitem [{\citenamefont {Cronenwett}\ \emph {et~al.}(1998)\citenamefont
  {Cronenwett}, \citenamefont {Oosterkamp},\ and\ \citenamefont
  {Kouwenhoven}}]{Cronenwett1998}%
  \BibitemOpen
  \bibfield  {author} {\bibinfo {author} {\bibfnamefont {S.~M.}\ \bibnamefont
  {Cronenwett}}, \bibinfo {author} {\bibfnamefont {T.~H.}\ \bibnamefont
  {Oosterkamp}}, \ and\ \bibinfo {author} {\bibfnamefont {L.~P.}\ \bibnamefont
  {Kouwenhoven}},\ }\href {http://science.sciencemag.org/content/281/5376/540}
  {\bibfield  {journal} {\bibinfo  {journal} {Science}\ }\textbf {\bibinfo
  {volume} {281}},\ \bibinfo {pages} {540} (\bibinfo {year}
  {1998})}\BibitemShut {NoStop}%
\bibitem [{\citenamefont {Longhi}(2016)}]{Longhi}%
  \BibitemOpen
  \bibfield  {author} {\bibinfo {author} {\bibfnamefont {S.}~\bibnamefont
  {Longhi}},\ }\href {https://link.aps.org/doi/10.1103/PhysRevA.93.022102}
  {\bibfield  {journal} {\bibinfo  {journal} {Phys. Rev. A}\ }\textbf {\bibinfo
  {volume} {93}},\ \bibinfo {pages} {022102} (\bibinfo {year}
  {2016})}\BibitemShut {NoStop}%
\bibitem [{\citenamefont {Anderson}(1970)}]{Anderson1970}%
  \BibitemOpen
  \bibfield  {author} {\bibinfo {author} {\bibfnamefont {P.~W.}\ \bibnamefont
  {Anderson}},\ }\href {http://stacks.iop.org/0022-3719/3/i=12/a=008}
  {\bibfield  {journal} {\bibinfo  {journal} {J. Phys. C: Solid State Phys.}\
  }\textbf {\bibinfo {volume} {3}},\ \bibinfo {pages} {2436} (\bibinfo {year}
  {1970})}\BibitemShut {NoStop}%
\bibitem [{\citenamefont {Wilson}(1975)}]{Wilson1975}%
  \BibitemOpen
  \bibfield  {author} {\bibinfo {author} {\bibfnamefont {K.~G.}\ \bibnamefont
  {Wilson}},\ }\href {\doibase 10.1103/RevModPhys.47.773} {\bibfield  {journal}
  {\bibinfo  {journal} {Rev. Mod. Phys.}\ }\textbf {\bibinfo {volume} {47}},\
  \bibinfo {pages} {773} (\bibinfo {year} {1975})}\BibitemShut {NoStop}%
\bibitem [{\citenamefont {Affleck}(1995)}]{Affleck:1995}%
  \BibitemOpen
  \bibfield  {author} {\bibinfo {author} {\bibfnamefont {I.}~\bibnamefont
  {Affleck}},\ }\href@noop {} {\bibfield  {journal} {\bibinfo  {journal} {Acta
  Phys. Polon.}\ }\textbf {\bibinfo {volume} {B26}},\ \bibinfo {pages} {1869}
  (\bibinfo {year} {1995})},\ \Eprint {http://arxiv.org/abs/cond-mat/9512099}
  {arXiv:cond-mat/9512099 [cond-mat]} \BibitemShut {NoStop}%
\bibitem [{\citenamefont {Chien}\ \emph {et~al.}(2015)\citenamefont {Chien},
  \citenamefont {Peotta},\ and\ \citenamefont {Di~Ventra}}]{Chien}%
  \BibitemOpen
  \bibfield  {author} {\bibinfo {author} {\bibfnamefont {C.-C.}\ \bibnamefont
  {Chien}}, \bibinfo {author} {\bibfnamefont {S.}~\bibnamefont {Peotta}}, \
  and\ \bibinfo {author} {\bibfnamefont {M.}~\bibnamefont {Di~Ventra}},\ }\href
  {\doibase 10.1038/nphys3531} {\bibfield  {journal} {\bibinfo  {journal} {Nat.
  Phys.}\ }\textbf {\bibinfo {volume} {11}},\ \bibinfo {pages} {998} (\bibinfo
  {year} {2015})}\BibitemShut {NoStop}%
\bibitem [{\citenamefont {Krinner}\ \emph {et~al.}(2017)\citenamefont
  {Krinner}, \citenamefont {Esslinger},\ and\ \citenamefont
  {Brantut}}]{Krinner2017}%
  \BibitemOpen
  \bibfield  {author} {\bibinfo {author} {\bibfnamefont {S.}~\bibnamefont
  {Krinner}}, \bibinfo {author} {\bibfnamefont {T.}~\bibnamefont {Esslinger}},
  \ and\ \bibinfo {author} {\bibfnamefont {J.-P.}\ \bibnamefont {Brantut}},\
  }\href {http://stacks.iop.org/0953-8984/29/i=34/a=343003} {\bibfield
  {journal} {\bibinfo  {journal} {J. Phys.: Condens. Matter}\ }\textbf
  {\bibinfo {volume} {29}},\ \bibinfo {pages} {343003} (\bibinfo {year}
  {2017})}\BibitemShut {NoStop}%
\bibitem [{\citenamefont {Robins}\ \emph {et~al.}(2008)\citenamefont {Robins},
  \citenamefont {Figl}, \citenamefont {Jeppesen}, \citenamefont {Dennis},\ and\
  \citenamefont {Close}}]{Robins2008}%
  \BibitemOpen
  \bibfield  {author} {\bibinfo {author} {\bibfnamefont {N.~P.}\ \bibnamefont
  {Robins}}, \bibinfo {author} {\bibfnamefont {C.}~\bibnamefont {Figl}},
  \bibinfo {author} {\bibfnamefont {M.}~\bibnamefont {Jeppesen}}, \bibinfo
  {author} {\bibfnamefont {G.~R.}\ \bibnamefont {Dennis}}, \ and\ \bibinfo
  {author} {\bibfnamefont {J.~D.}\ \bibnamefont {Close}},\ }\href
  {http://dx.doi.org/10.1038/nphys1027} {\bibfield  {journal} {\bibinfo
  {journal} {Nat. Phys.}\ }\textbf {\bibinfo {volume} {4}},\ \bibinfo {pages}
  {731 EP } (\bibinfo {year} {2008})}\BibitemShut {NoStop}%
\bibitem [{\citenamefont {Esslinger}(2010)}]{Esslinger2010}%
  \BibitemOpen
  \bibfield  {author} {\bibinfo {author} {\bibfnamefont {T.}~\bibnamefont
  {Esslinger}},\ }\href
  {https://doi.org/10.1146/annurev-conmatphys-070909-104059} {\bibfield
  {journal} {\bibinfo  {journal} {Annu. Rev. Condens. Matter Phys.}\ }\textbf
  {\bibinfo {volume} {1}},\ \bibinfo {pages} {129} (\bibinfo {year}
  {2010})}\BibitemShut {NoStop}%
\bibitem [{\citenamefont {Riegger}\ \emph {et~al.}(2018)\citenamefont
  {Riegger}, \citenamefont {Darkwah~Oppong}, \citenamefont {H\"ofer},
  \citenamefont {Fernandes}, \citenamefont {Bloch},\ and\ \citenamefont
  {F\"olling}}]{Riegger2018}%
  \BibitemOpen
  \bibfield  {author} {\bibinfo {author} {\bibfnamefont {L.}~\bibnamefont
  {Riegger}}, \bibinfo {author} {\bibfnamefont {N.}~\bibnamefont
  {Darkwah~Oppong}}, \bibinfo {author} {\bibfnamefont {M.}~\bibnamefont
  {H\"ofer}}, \bibinfo {author} {\bibfnamefont {D.~R.}\ \bibnamefont
  {Fernandes}}, \bibinfo {author} {\bibfnamefont {I.}~\bibnamefont {Bloch}}, \
  and\ \bibinfo {author} {\bibfnamefont {S.}~\bibnamefont {F\"olling}},\ }\href
  {https://link.aps.org/doi/10.1103/PhysRevLett.120.143601} {\bibfield
  {journal} {\bibinfo  {journal} {Phys. Rev. Lett.}\ }\textbf {\bibinfo
  {volume} {120}},\ \bibinfo {pages} {143601} (\bibinfo {year}
  {2018})}\BibitemShut {NoStop}%
\bibitem [{\citenamefont {Simon}\ and\ \citenamefont {Affleck}(2001)}]{Simon}%
  \BibitemOpen
  \bibfield  {author} {\bibinfo {author} {\bibfnamefont {P.}~\bibnamefont
  {Simon}}\ and\ \bibinfo {author} {\bibfnamefont {I.}~\bibnamefont
  {Affleck}},\ }\href {\doibase 10.1103/PhysRevB.64.085308} {\bibfield
  {journal} {\bibinfo  {journal} {Phys. Rev. B}\ }\textbf {\bibinfo {volume}
  {64}},\ \bibinfo {pages} {085308} (\bibinfo {year} {2001})}\BibitemShut
  {NoStop}%
\bibitem [{\citenamefont {{Nozi\`eres, Ph.}}\ and\ \citenamefont {{Blandin,
  A.}}(1980)}]{Nozieresph}%
  \BibitemOpen
  \bibfield  {author} {\bibinfo {author} {\bibnamefont {{Nozi\`eres, Ph.}}}\
  and\ \bibinfo {author} {\bibnamefont {{Blandin, A.}}},\ }\href
  {https://doi.org/10.1051/jphys:01980004103019300} {\bibfield  {journal}
  {\bibinfo  {journal} {J. Phys. France}\ }\textbf {\bibinfo {volume} {41}},\
  \bibinfo {pages} {193} (\bibinfo {year} {1980})}\BibitemShut {NoStop}%
\bibitem [{\citenamefont {Zawadowski}(1980)}]{Zawadowski}%
  \BibitemOpen
  \bibfield  {author} {\bibinfo {author} {\bibfnamefont {A.}~\bibnamefont
  {Zawadowski}},\ }\href {\doibase 10.1103/PhysRevLett.45.211} {\bibfield
  {journal} {\bibinfo  {journal} {Phys. Rev. Lett.}\ }\textbf {\bibinfo
  {volume} {45}},\ \bibinfo {pages} {211} (\bibinfo {year} {1980})}\BibitemShut
  {NoStop}%
\bibitem [{\citenamefont {Nozi{\`e}res}(1974)}]{Nozieres1974}%
  \BibitemOpen
  \bibfield  {author} {\bibinfo {author} {\bibfnamefont {P.}~\bibnamefont
  {Nozi{\`e}res}},\ }\href {https://doi.org/10.1007/BF00654541} {\bibfield
  {journal} {\bibinfo  {journal} {J. Low Temp. Phys.}\ }\textbf {\bibinfo
  {volume} {17}},\ \bibinfo {pages} {31} (\bibinfo {year} {1974})}\BibitemShut
  {NoStop}%
\bibitem [{\citenamefont {{Glazman}}\ and\ \citenamefont
  {{Pustilnik}}(2003)}]{Glazman2003}%
  \BibitemOpen
  \bibfield  {author} {\bibinfo {author} {\bibfnamefont {L.~I.}\ \bibnamefont
  {{Glazman}}}\ and\ \bibinfo {author} {\bibfnamefont {M.}~\bibnamefont
  {{Pustilnik}}},\ }\href@noop {} {\bibfield  {journal} {\bibinfo  {journal}
  {eprint arXiv:cond-mat/0302159}\ } (\bibinfo {year} {2003})},\ \Eprint
  {http://arxiv.org/abs/cond-mat/0302159} {cond-mat/0302159} \BibitemShut
  {NoStop}%
\bibitem [{\citenamefont {Datta}(1997)}]{Datta1997}%
  \BibitemOpen
  \bibfield  {author} {\bibinfo {author} {\bibfnamefont {S.}~\bibnamefont
  {Datta}},\ }\href@noop {} {\emph {\bibinfo {title} {Electronic Transport in
  Mesoscopic Systems}}}\ (\bibinfo  {publisher} {Cambridge University Press},\
  \bibinfo {year} {1997})\BibitemShut {NoStop}%
\bibitem [{\citenamefont {Pustilnik}\ and\ \citenamefont
  {Glazman}(2001)}]{Pustilnik}%
  \BibitemOpen
  \bibfield  {author} {\bibinfo {author} {\bibfnamefont {M.}~\bibnamefont
  {Pustilnik}}\ and\ \bibinfo {author} {\bibfnamefont {L.~I.}\ \bibnamefont
  {Glazman}},\ }\href {https://link.aps.org/doi/10.1103/PhysRevB.64.045328}
  {\bibfield  {journal} {\bibinfo  {journal} {Phys. Rev. B}\ }\textbf {\bibinfo
  {volume} {64}},\ \bibinfo {pages} {045328} (\bibinfo {year}
  {2001})}\BibitemShut {NoStop}%
\bibitem [{\citenamefont {Ng}\ and\ \citenamefont {Lee}(1988)}]{Tai}%
  \BibitemOpen
  \bibfield  {author} {\bibinfo {author} {\bibfnamefont {T.~K.}\ \bibnamefont
  {Ng}}\ and\ \bibinfo {author} {\bibfnamefont {P.~A.}\ \bibnamefont {Lee}},\
  }\href {\doibase 10.1103/PhysRevLett.61.1768} {\bibfield  {journal} {\bibinfo
   {journal} {Phys. Rev. Lett.}\ }\textbf {\bibinfo {volume} {61}},\ \bibinfo
  {pages} {1768} (\bibinfo {year} {1988})}\BibitemShut {NoStop}%
\bibitem [{\citenamefont {Affleck}\ and\ \citenamefont
  {Ludwig}(1991)}]{AffleckLudwig}%
  \BibitemOpen
  \bibfield  {author} {\bibinfo {author} {\bibfnamefont {I.}~\bibnamefont
  {Affleck}}\ and\ \bibinfo {author} {\bibfnamefont {A.~W.}\ \bibnamefont
  {Ludwig}},\ }\href
  {http://www.sciencedirect.com/science/article/pii/055032139190419X}
  {\bibfield  {journal} {\bibinfo  {journal} {Nucl. Phys. B}\ }\textbf
  {\bibinfo {volume} {360}},\ \bibinfo {pages} {641 } (\bibinfo {year}
  {1991})}\BibitemShut {NoStop}%
\bibitem [{\citenamefont {Lee}\ and\ \citenamefont {Toner}(1992)}]{Lee1992}%
  \BibitemOpen
  \bibfield  {author} {\bibinfo {author} {\bibfnamefont {D.-H.}\ \bibnamefont
  {Lee}}\ and\ \bibinfo {author} {\bibfnamefont {J.}~\bibnamefont {Toner}},\
  }\href {https://link.aps.org/doi/10.1103/PhysRevLett.69.3378} {\bibfield
  {journal} {\bibinfo  {journal} {Phys. Rev. Lett.}\ }\textbf {\bibinfo
  {volume} {69}},\ \bibinfo {pages} {3378} (\bibinfo {year}
  {1992})}\BibitemShut {NoStop}%
\bibitem [{\citenamefont {Furusaki}\ and\ \citenamefont
  {Nagaosa}(1994)}]{Furusaki1994}%
  \BibitemOpen
  \bibfield  {author} {\bibinfo {author} {\bibfnamefont {A.}~\bibnamefont
  {Furusaki}}\ and\ \bibinfo {author} {\bibfnamefont {N.}~\bibnamefont
  {Nagaosa}},\ }\href {\doibase 10.1103/PhysRevLett.72.892} {\bibfield
  {journal} {\bibinfo  {journal} {Phys. Rev. Lett.}\ }\textbf {\bibinfo
  {volume} {72}},\ \bibinfo {pages} {892} (\bibinfo {year} {1994})}\BibitemShut
  {NoStop}%
\bibitem [{\citenamefont {Lodahl}\ \emph {et~al.}(2017)\citenamefont {Lodahl},
  \citenamefont {Mahmoodian}, \citenamefont {Stobbe}, \citenamefont
  {Rauschenbeutel}, \citenamefont {Schneeweiss}, \citenamefont {Volz},
  \citenamefont {Pichler},\ and\ \citenamefont {Zoller}}]{Lodahl}%
  \BibitemOpen
  \bibfield  {author} {\bibinfo {author} {\bibfnamefont {P.}~\bibnamefont
  {Lodahl}}, \bibinfo {author} {\bibfnamefont {S.}~\bibnamefont {Mahmoodian}},
  \bibinfo {author} {\bibfnamefont {S.}~\bibnamefont {Stobbe}}, \bibinfo
  {author} {\bibfnamefont {A.}~\bibnamefont {Rauschenbeutel}}, \bibinfo
  {author} {\bibfnamefont {P.}~\bibnamefont {Schneeweiss}}, \bibinfo {author}
  {\bibfnamefont {J.}~\bibnamefont {Volz}}, \bibinfo {author} {\bibfnamefont
  {H.}~\bibnamefont {Pichler}}, \ and\ \bibinfo {author} {\bibfnamefont
  {P.}~\bibnamefont {Zoller}},\ }\href {http://dx.doi.org/10.1038/nature21037}
  {\bibfield  {journal} {\bibinfo  {journal} {Nature}\ }\textbf {\bibinfo
  {volume} {541}},\ \bibinfo {pages} {473} (\bibinfo {year}
  {2017})}\BibitemShut {NoStop}%
\bibitem [{\citenamefont {Bulla}\ \emph {et~al.}(2008)\citenamefont {Bulla},
  \citenamefont {Costi},\ and\ \citenamefont {Pruschke}}]{Bulla2008}%
  \BibitemOpen
  \bibfield  {author} {\bibinfo {author} {\bibfnamefont {R.}~\bibnamefont
  {Bulla}}, \bibinfo {author} {\bibfnamefont {T.~A.}\ \bibnamefont {Costi}}, \
  and\ \bibinfo {author} {\bibfnamefont {T.}~\bibnamefont {Pruschke}},\ }\href
  {https://link.aps.org/doi/10.1103/RevModPhys.80.395} {\bibfield  {journal}
  {\bibinfo  {journal} {Rev. Mod. Phys.}\ }\textbf {\bibinfo {volume} {80}},\
  \bibinfo {pages} {395} (\bibinfo {year} {2008})}\BibitemShut {NoStop}%
\end{thebibliography}%

\end{document}